\documentclass[aps,preprint,superscriptaddress]{revtex4-1}

\usepackage{graphicx}
\usepackage{amsmath}
\usepackage{amssymb}
\usepackage{slashed}

\newcommand{\tr}{\rm tr \,}

\usepackage{array} 
\newcolumntype{C}{>{$}c<{$}}

\begin{document}


\title{Low-energy constants in the chiral Lagrangian \\ with baryon octet and decuplet fields \\ from Lattice QCD data on CLS ensembles}



\author{Matthias F.M. Lutz}
\affiliation{GSI Helmholtzzentrum f\"ur Schwerionenforschung GmbH, \\Planckstra\ss e 1, 64291 Darmstadt, Germany}
\author{Yonggoo Heo}
\affiliation{Bogoliubov Laboratory for Theoretical Physics, Joint Institute for Nuclear Research, RU-141980 Dubna, Moscow region, Russia}
\affiliation{GSI Helmholtzzentrum f\"ur Schwerionenforschung GmbH, \\Planckstra\ss e 1, 64291 Darmstadt, Germany}
\author{Xiao-Yu Guo}
\affiliation{Faculty of Science, Beijing University of Technology,\\
  Beijing 100124, China}
\date{\today}

\begin{abstract}

We perform an analysis of Lattice QCD data on baryon octet and decuplet masses based on the chiral SU(3) Lagrangian. Low-energy constants (LEC) are adjusted to describe baryon masses from a large set of CLS ensembles, where finite-box and discretization effects are considered.  The set is successfully compared against previous Lattice QCD data from ensembles generated with distinct QCD actions by the ETMC, QCDSF-UKQCD and HSC groups. 
Discretization effects are modelled by the use of action and lattice-scale dependent leading orders LEC, where uniform values are imposed in the limit of vanishing lattice scales. 
From the CLS 
data set we extract a pion-nucleon sigma term, $\sigma_{\pi N}= 58.7(1.2)$ MeV,  compatible with its empirical value and a sizeable strangeness content of the nucleon with $\sigma_{sN} = -316(76)$ MeV. 

\end{abstract}


\maketitle
\tableofcontents

\newpage


\section{Introduction}
\label{sec:1}

An improved insight into the quark-mass dependence of hadron masses is the key element for the construction of a theory bridge from QCD towards scattering data in the laboratory. 
This point is made explicit by the chiral Lagrangian properly constructed for some heavy hadron, in which interaction terms that are relevant in a computation of the hadron mass, play a decisive role also in the scattering of the hadron off a pion, kaon or eta \cite{Guo:2018kno,Lutz:2022enz,Korpa:2022voo}. 

Such a strategy was recently scrutinized by the authors in the open-charm sector of QCD, for which there exists a sizable data base from Lattice QCD simulation by ETMC, HPQCD and HSC \cite{Kalinowski:2015bwa,Na:2012iu,Moir:2016srx,Cheung:2020mql,Gayer:2021xzv}. A global fit 
of the Low-Energy Constants (LEC) to the D meson masses and to some s- and p-wave scattering phase shifts was obtained \cite{Guo:2018kno,Lutz:2022enz}.  While such a system serves as a useful first test bed for our strategy, it comes with a rather scarce empirical data set only. It is hard to extract information on the scattering of a pion, kaon or eta off the D mesons from accelerator experiments. Here the baryon masses composed out of up, down and strange quarks are of more importance,
simply because there is already a large experimental data base on pion-nucleon and kaon-nucleon scattering collected over the last few decades that still awaits a profound theory reproduction from QCD dynamics. 

Our work  is triggered by the availability of a new and large set of baryon masses computed on CLS Lattice QCD ensembles by the Regensburg group \cite{RQCD:2022xux}. Results are shown for six distinct values of the QCD beta value that allow a continuum limit extrapolation of the finite box baryon masses. We will analyze this data set with extrapolation framework that was derived from the chiral Lagrangian at the one-loop level, where loop contributions are evaluated in terms of on-shell meson and baryon masses. In our previous studies \cite{Lutz:2012mq,Lutz:2014oxa,Lutz:2018cqo,Guo:2019nyp} a successful global reproduction of baryon masses on ensembles by 
PACS-CS, LHPC, HSC, NPLQCD, QCDSF-UKQCD and ETMC \cite{PACS-CS:2008bkb,HadronSpectrum:2008xlg,Walker-Loud:2008rui,Beane:2011pc,Bietenholz:2011qq,Alexandrou:2013joa} was achieved in terms of a set of LEC. So far, with the older data set, a full continuum extrapolation of baryon masses was not possible. Therefore our previous LEC may be subject to important changes. 

The work is organized as follows. In Section II the terms relevant in the computation of the baryon masses are presented in combination of a set of sum rules that arise from QCD at a large number of colors \cite{Lutz:2010se,Lutz:2018cqo}. It follows in Section III  a primer on the set of nonlinear and coupled equations that determine the meson and baryon masses in our extrapolation framework \cite{Lutz:2018cqo,Guo:2019nyp}. 
Results from a fit to the Lattice QCD data set are discussed then in Section IV. We close with a summary and conclusions.

\section{Low-energy constants in the chiral Lagrangian}
\label{sec:1}

We recall the set of Low Energy Constants (LEC) that turn relevant in a computation of the baryon octet and decuplet masses at next-to-next-to-next-to-leading order (N$^3$LO) from the chiral Lagrangian  \cite{Gasser:1987rb,Jenkins:1990jv,Krause:1990xc,Becher:1999he,Lutz:2001yb,Scherer:2002tk,Lehnhart:2004vi,Semke:2005sn,Hacker:2005fh,Frink:2006hx,MartinCamalich:2010fp,Semke:2011ez,Semke:2012gs,Lutz:2012mq,Ren:2012aj,Ren:2013oaa,Lutz:2014oxa,Lutz:2018cqo,Holmberg:2018dtv}. The applied conventions are illustrated with the leading order terms 
\begin{eqnarray}
&& \mathcal{L}^{(1)} =
\mathrm{tr}\, \Big\{ \bar B\, (i\, D^\alpha \,\gamma_\alpha\, - M_{[8]})\, B \Big\}  -
 \mathrm{tr}\, \Big\{ \bar B_\mu \cdot \big((i\,D^\alpha \gamma_\alpha\, - M_{[10]})\,g^{\mu\nu} 
\nonumber\\ 
&& \qquad \qquad \qquad  -\,i\,(\gamma^\mu D^\nu + \gamma^\nu D^\mu) + \gamma^\mu(i\,D^\alpha \gamma_\alpha + M_{[10]})\gamma^\nu \big)\, B_\nu \Big\}\,,
\nonumber\\
&& \qquad + F\, \mathrm{tr} \Big\{ \bar{B}\, \gamma^\mu \gamma_5\, [i\,U_\mu,B]\, \Big\} + D\, \mathrm{tr}\Big\{ \bar{B}\, \gamma^\mu \gamma_5\, \{i\,U_\mu,\,B\}\, \Big\}
\nonumber \\
&& \qquad +\, C\left( \mathrm{tr} \Big\{ (\bar{B}_\mu \cdot i\, U^\mu)\, B\Big\} + \mathrm{h.c.} \right)
+ H\, \mathrm{tr} \Big\{ (\bar{B}^\mu\cdot \gamma_5\,\gamma_\nu   B_\mu)\, i\,U^\nu \Big\}\,,
\nonumber\\ \nonumber\\
&&\Gamma_\mu ={\textstyle{1\over 2}}\,u^\dagger \,\big(\partial_\mu \,u \big)
+{\textstyle{1\over 2}}\, u \,
\big(\partial_\mu\,u^\dagger  \big) \,,  \qquad \qquad
U_\mu = {\textstyle{1\over 2}}\,u^\dagger \, \big(
\partial_\mu \,e^{i\,\frac{\Phi}{f}} \big)\, u^\dagger\;, \qquad \qquad
 u = e^{i\,\frac{\Phi}{2\,f}} \,,
\nonumber\\
&&  D_\mu \, B \;\,= \partial_\mu B +  \Gamma_{\mu}\, B -
B\,\Gamma_{\mu} \,, 
\label{def-L1}
\end{eqnarray}
for the baryon spin 1/2 and 3/2 fields, $B$ and $B_\mu$. The kinetic terms in (\ref{def-L1}) 
encounter the covariant derivative $D_\mu$ and  the SU(3) matrix field of the Goldstone bosons $\Phi$. We apply the convenient flavor '$\cdot $' product notation for terms involving the symmetric baryon fields $B_\mu$ and $\bar B_\mu$ (see  \cite{Lutz:2001yb,Semke:2011ez}). For instance, the product of the two flavor decuplet fields  $\bar B_\mu \cdot B_\nu $  is constructed
to transform as a flavor octet.  At this order there are seven LEC encountered, the chiral limit values of the baryon masses, axial-vector coupling constants, $M_{[8]}, M_{[10]}$ and
$F, D, C, H$ together with the meson decay parameter $f$.

The Goldstone boson sector of the Lagrangian  \cite{Gasser:1984gg}  is well established with
\begin{eqnarray}
&&\mathcal{L}^{(2)} = -\, f^2\,{\tr } U_\mu\,U^\mu +\frac{1}{2}\,f^2\,{\tr } \chi_+ \,,
\nonumber\\
&& \mathcal{L}^{(4)}  =  16\,L_1\,({\tr}U_\mu\,U^\mu)^2 + 16\,L_2\,{\tr} U_\mu\,U_\nu\,{\tr}U^\mu\,U^\nu 
+ 16\,L_3\,{\tr}U_\mu\,U^\mu\,U_\nu\,U^\nu 
\nonumber\\
&& \qquad \, - \,8\,L_4\,{\tr}U_\mu\,U^\mu\,{\tr}\chi_+- 8\,L_5\,{\tr}U_\mu\,U^\mu\,\chi_+ 
+ 4\,L_6\,({\tr}\chi_+)^2
\nonumber\\
&& \qquad \,+\, 4\,L_7\,({\tr}\chi_-)^2
+2\,L_8\,{\tr}(\chi_+ \chi_+ + \chi_- \chi_- )\,,
\nonumber\\
&& \chi_\pm = {\textstyle \frac{1}{2}} \left(
u\,\chi_0 \,u
\pm u^\dagger\,\chi_0 \,u^\dagger 
\right) \,, \qquad \qquad \chi_0 =2\,B_0\, {\rm diag} (m_u,m_d,m_s) \,,
\label{def-L42}
\end{eqnarray}
in terms of the eight LEC of Gasser and Leutwyler $L_{1-8}$. The terms in (\ref{def-L42}) play an instrumental role in the translation of the quark-mass parameters to the masses of the pseudo-Goldstone bosons as measured on various Lattice QCD ensembles (see e.g. \cite{Lutz:2018cqo,Guo:2018kno,Bavontaweepanya:2018yds,Guo:2018zvl,Aoki:2021kgd}). Here the chiral symmetry breaking fields $\chi_\pm$ proportional to the quark masses of QCD are encountered.

The next-to-leading terms in the baryon part of the chiral Lagrangian that contribute to the baryon masses at the one-loop level are
\allowdisplaybreaks[1]
\begin{eqnarray}
&& \mathcal{L}^{(2)}_\chi = 2\, b_0 \,\mathrm{tr} \left(\bar B \,B\right) \mathrm{tr}\left(\chi_+\right) + 2 \,b_D\,\mathrm{tr}\left(\bar{B}\,\{\chi_+,\,B\}\right) + 2\, b_F\,\mathrm{tr}\left(\bar{B}\,[\chi_+,\,B]\right) \nonumber \\
&& \qquad  -\, 2\, d_0\, \mathrm{tr}\left(\bar B_\mu \cdot B^\mu\right) \mathrm{tr}(\chi_+) - 2\, d_D\, \mathrm{tr} \left( \left(\bar{B}_\mu \cdot B^\mu\right) \chi_+\right)\,,
\nonumber\\
&& \mathcal{L}^{(2)}_S =- \frac{1}{2}\,g_0^{(S)}\,\mathrm{tr} \,\Big\{\bar{B}\,B \Big\}\, \mathrm{tr}\Big\{ U_\mu\, U^\mu \Big\} - \frac{1}{2}\,g_1^{(S)}\,\mathrm{tr} \,\Big\{ \bar{B}\, U^\mu \Big\}\, \mathrm{tr}\,\Big\{U_\mu\, B \Big\}
\nonumber \\
&&\qquad -\,\frac{1}{4}\,g_D^{(S)} \mathrm{tr}\,\Big\{\bar{B}\left\{\left\{U_\mu, U^\mu\right\}, B\right\}\Big\}
-\frac{1}{4}\,g_F^{(S)}\mathrm{tr}\,\Big\{ \bar{B}\left[\left\{U_\mu, U^\mu\right\}, B\right]\Big\}
\nonumber\\
&&\qquad + \,\frac{1}{2}\,h_1^{(S)}\,\mathrm{tr}\,\Big\{ \bar{B}_\mu \cdot B^\mu \Big\}\, \mathrm{tr}\,\Big\{U_\nu\; U^\nu\Big\} +
\frac{1}{2}\,h_2^{(S)}\,\mathrm{tr}\,\Big\{\bar{B}_\mu \cdot B^\nu \Big\}\, \mathrm{tr}\,\Big\{U^\mu\, U_\nu\Big\}
\nonumber \\
&& \qquad + \,h_3^{(S)}\,\mathrm{tr}\,\Big\{\big(\bar{B}_\mu \cdot B^\mu\big)\, \big(U^\nu\, U_\nu\big) \Big\} + \frac{1}{2}\,h_4^{(S)}\,\mathrm{tr}\,\Big\{ \big(\bar{B}_\mu \cdot B^\nu\big)\, \{U^\mu,\, U_\nu \} \Big\}
\nonumber \\
&&\qquad  +\, h_5^{(S)}\, \mathrm{tr}\, \Big\{ \big( \bar{B}_\mu \cdot U_\nu\big)\, \big(U^\nu\cdot B^\mu \big) \Big\}
\nonumber \\
&& \qquad +\, \frac{1}{2}\,h_6^{(S)}\, \mathrm{tr} \Big\{ \big( \bar{B}_\mu \cdot U^\mu\big)\, \big(U^\nu\cdot B_\nu \big)
+\big( \bar{B}_\mu \cdot U^\nu\big)\, \big(U^\mu\cdot B_\nu \big) \Big\} \, ,
\nonumber\\ 
&&\mathcal{L}^{(2)}_V = -\frac{1}{4}\,g_0^{(V)}\, \Big( \mathrm{tr}\,\Big\{\bar{B}\, i\,\gamma^\mu\, D^\nu B\Big\} \,
\mathrm{tr}\,\Big\{ U_\nu\, U_\mu \Big\}\Big)
\nonumber \\
&& \qquad -\, \frac{1}{8}\,g_1^{(V)} \,\Big( \mathrm{tr}\,\Big\{\bar{B}\,U_\mu \Big\} \,i\,\gamma^\mu \, \mathrm{tr}\,\Big\{U_\nu\, D^\nu B\Big\} 
+ \mathrm{tr}\,\Big\{\bar{B}\,U_\nu \Big\} \,i\,\gamma^\mu \, \mathrm{tr}\,\Big\{U_\mu\, D^\nu B\Big\} \Big)
\nonumber \\
&& \qquad -\, \frac{1}{8}\,g_D^{(V)}\, \Big(\mathrm{tr}\,\Big\{\bar{B}\, i\,\gamma^\mu \left\{\left\{U_\mu,\, U_\nu\right\}, D^\nu B\right\}\Big\} \Big)
\nonumber\\
&& \qquad -\, \frac{1}{8}\,g_F^{(V)}\,\Big( \mathrm{tr}\,\Big\{ \bar{B}\, i\,\gamma^\mu\, \left[\left\{U_\mu,\, U_\nu\right\},\, D^\nu B \right]\Big\}  \Big)
\nonumber \\
&& \qquad +\, \frac{1}{4}\,h_1^{(V)}\,\Big(\mathrm{tr}\,\Big\{ \bar{B}_\lambda \cdot i\,\gamma^\mu\, D^\nu B^\lambda\Big\} \,\mathrm{tr}\,\Big\{U_\mu\, U_\nu\Big\}\Big)
\nonumber \\
&& \qquad +\, \frac{1}{4}\,h_2^{(V)}\,\Big(\mathrm{tr}\,\Big\{ \left(\bar{B}_\lambda \cdot i\,\gamma^\mu\, D^\nu B^\lambda \right) \{U_\mu,\, U_\nu\}\Big\} \Big)
\nonumber \\
&& \qquad +\, \frac{1}{4}\,h_3^{(V)}\, \Big( \mathrm{tr}\, \Big\{ \left( \bar{B}_\lambda \cdot U_\mu\right) i\,\gamma^\mu \left(U_\nu\cdot D^\nu B^\lambda \right)
\nonumber\\
&& \qquad \qquad \qquad \qquad +\, \left( \bar{B}_\lambda \cdot U_\nu\right) i\,\gamma^\mu \left(U_\mu\cdot D^\nu B^\lambda \right) \Big\} \Big)
 + \mathrm{h.c.} \, ,
\label{def-Q2-terms}
\end{eqnarray}
where we emphasize the dual role played by the LEC in (\ref{def-Q2-terms}). On the one-hand such terms contribute to the baryon masses, on the other hand, they predict the next-to-leading-order tree-level contributions to the meson-baryon scattering processes. That is why a dedicated study of the quark-mass dependence of the baryon masses generates important insight into the dynamics of such scattering processes. At N$^3$LO two further sets of LEC are needed:
\begin{eqnarray}
&& \mathcal{L}^{(3)}_\chi = \zeta_0\, \mathrm{tr} \big(\bar{B}\, (i\,( D^\alpha \gamma_\alpha ) -M_{[8]})\, B\big)\, \mathrm{tr}(\chi_+) + \zeta_D\, \mathrm{tr} \big(\bar{B}\, (i\,( D^\alpha \gamma_\alpha ) -M_{[8]})\, \{\chi_+,B \} \big) \nonumber\\
&& \qquad + \,\zeta_F\, \mathrm{tr} \big(\bar{B}\, (i\,( D^\alpha \gamma_\alpha ) -M_{[8]})\, [ \chi_+ ,\,B] \big) 
- \xi^{}_0\, \mathrm{tr} \big(\bar B_\mu\,\cdot  (i\,( D^\alpha \gamma_\alpha ) -M_{[10]})\, B^\mu \big)\, \mathrm{tr}(\chi_+) 
\nonumber \\
&& \qquad -\, \xi^{}_D\, \mathrm{tr} \big( \big(\bar B_\mu\,\cdot (i\,( D^\alpha \gamma_\alpha ) -M_{[10]})\,B^\mu \big)\, \chi_+ \big) \,,
\nonumber\\ 
&& \mathcal{L}^{(4)}_\chi = c_0\, \mathrm{tr}\left(\bar B\, B\right) \mathrm{tr} \left(\chi_+^2\right) + c_1\, \mathrm{tr} \left(\bar B \,\chi_+\right) \mathrm{tr}\left(\chi_+\, B\right) 
\nonumber \\
&& \qquad + c_2\, \mathrm{tr} \left( \bar B\, \{\chi_+^2,\, B\} \right)  + c_3\,\mathrm{tr} \left( \bar B\, [\chi_+^2, \,B] \right)
\nonumber \\
&&  \qquad +c_4\, \mathrm{tr} \left(\bar B\, \{\chi_+,\,B\} \right) \mathrm{tr} (\chi_+) + c_5\, \mathrm{tr}\left(\bar B\, [\chi_+,\,B]\right) \mathrm{tr} (\chi_+) \nonumber \\
&& \qquad + c_6\, \mathrm{tr} \left(\bar B \,B\right) \left(\mathrm{tr}(\chi_+)\right)^2
\nonumber \\
&&  \qquad -e_0\, \mathrm{tr}\left(\bar B_\mu \cdot B^\mu \right) \mathrm{tr}\left(\chi_+^2\right) - e_1\, \mathrm{tr}\left( \left(\bar{B}_\mu \cdot \chi_+\right) \left(\chi_+ \cdot B^\mu\right) \right)
\nonumber \\
&& \qquad - e_2\, \mathrm{tr}\left( \left(\bar B_\mu \cdot B^\mu\right)\chi_+^2\right) - e_3\, \mathrm{tr}\left( \left(\bar B_\mu \cdot B^\mu\right)\chi_+\right) \mathrm{tr}(\chi_+) \nonumber \\
&&  \qquad-e_4\, \mathrm{tr}\left(\bar B_\mu \cdot B^\mu\right) \left(\mathrm{tr}(\chi_+)\right)^2\,,
\label{def-c-e}
\end{eqnarray}
where again we focus on terms that contribute to the baryon masses. 

Altogether we identified $7+5+10+7 +5 + 12= 46$  LEC in the baryon sector. Like in our previous works we impose a set of sum rules that were derived from QCD at large numbers of colors ($N_c$) \cite{tHooft:1973alw,Witten:1979kh,Dashen:1993jt,Luty:1993fu,Lutz:2010se,Lutz:2018cqo}. 
The 14 sum rules follow at subleading order in that expansion
\begin{eqnarray}
&&  C=2\,D\,,\qquad H= 9\,F-3\,D \,, \qquad  
\nonumber\\ 
&&  \bar g^{(S)}_F =   \bar g^{(S)}_0 + \tfrac{3}{2}\,  \bar g^{(S)}_1 +   \bar g^{(S)}_D -   \bar h^{(S)}_1 - \tfrac{1}{3}\,  \bar h^{(S)}_5 + \tfrac{2}{9}\,  \bar h^{(S)}_6 \,,
\nonumber\\
&&   \bar h^{(S)}_2 = 0\,, \qquad   \bar h^{(S)}_4 = -   \bar h^{(S)}_6\,,
\nonumber\\
&&   \bar h^{(S)}_3 = \tfrac{3}{2}\, \bar g^{(S)}_0 + \tfrac{15}{4}\,  \bar g^{(S)}_1 + 3\,  \bar g^{(S)}_D - \tfrac{3}{2}\,  \bar h^{(S)}_1 
- \tfrac{3}{2}\,  \bar h^{(S)}_5 + \tfrac{1}{3}\,  \bar h^{(S)}_6 \,, \qquad 
\nonumber\\   
&&  \bar g^{(V)}_F =   \bar g^{(V)}_0 + \tfrac{3}{2}\,  \bar g^{(V)}_1 +   \bar g^{(V)}_D -   \bar h^{(V)}_1 - \tfrac{1}{3}\,  \bar h^{(V)}_3\,, \qquad
\nonumber\\      
&&    \bar h^{(V)}_2 = \tfrac{3}{2}\,  \bar g^{(V)}_0 + \tfrac{15}{4}\,  \bar g^{(V)}_1 + 3\,  \bar g^{(V)}_D - \tfrac{3}{2}\,  \bar h^{(V)}_1 - \tfrac{3}{2}\,  \bar h^{(V)}_3\,,
\label{Q4-subleading}\nonumber\\ 
&& 3\,\zeta_F + 3\,\zeta_D= \xi_D \,,\qquad \qquad  
\xi_0+ \tfrac{1}{3}\,\xi_D=\zeta_0+ 2\,\zeta_D\,,
\nonumber\\  
&& c_0 = 2\,c_3 + e_0 - {\textstyle{ 1\over 6}}\,e_1 - {\textstyle{ 1\over 3}}\,e_2 - {\textstyle{ 1\over 2}}\,c_1\,, \qquad 
c_1 = {\textstyle{ 1\over 3}}\,(e_1 + e_2) - c_2 - c_3\,,
\nonumber\\
&& e_3 = 3\,(c_4 + c_5)\,, \qquad
 e_4 = c_0 + c_2 + c_4 + c_6 - c_3 - c_5 - e_0\,,
\label{res-LargeNc}
\end{eqnarray}
where we impose such relations on LEC only, that affect the baryon masses at N$^2$LO or  N$^3$LO. That leaves 32 LEC to be determined from Lattice QCD data on the baryon masses. 
According to \cite{Lutz:2018cqo} the sum rules for the scalar and vector two-body LEC $g^{(S,V)}_{\cdots }$ and $h^{(S,V)}_{\cdots }$ hold for properly renormalized 'bar' values as recalled in the Appendix.

It is convenient to fix further 8 LEC by the requirement that at physical quark masses the empirical isospin-averaged 
baryon masses with $m_u=m_d = m$  are recovered in our approach. 
Our fits to the lattice QCD data sets involve 24 independent LEC of the chiral Lagrangian in its baryon sector. Given the large number of data points, of about 400, from the baryon masses on the CLS ensembles this appears to be a reasonable strategy.

\section{Nonlinear systems for meson and baryon masses}
\label{sec:2}

We turn to the loop contributions. Since we depart from the conventional Chiral Perturbation Theory ($\chi$PT) approach, we illustrate our method \cite{Lutz:2018cqo} at the hand of the meson masses first. For given quark masses, $m_u = m_d = m$ and $m_s$, the pion, kaon and eta masses are determined as the solution of a set of three coupled and nonlinear equations of the following form
\allowdisplaybreaks[1]
\begin{eqnarray}
&& m_\pi^2 \,=2\,B_0\,m - \frac{1}{18\,f^2}\,\Big\{-10\, m_\pi^2 + \,4\,m_K^2- 3\,m_\eta^2 \Big\}\,\bar I_\pi  - \frac{1}{6\,f^2}\,m_\pi^2\,\bar I_\eta 
\nonumber\\
&& \quad \; \;\; +\,\frac{8}{f^2}\,m_\pi^2\,(m_\pi^2 + \,2\,m_K^2)\,(2\,L_6-L_4) 
+\,\frac{8}{f^2}\,m_\pi^4\, (2\,L_8-L_5)\,,
\nonumber\\
&& m_K^2 = B_0\,(m+m_s) -\,\frac{1}{6\,f^2}\,\Big\{ m_\pi^2 -\,4\, m_K^2 +\,3\,m_\eta^2 \Big\}\,\bar I_K+\,\frac{1}{3\,f^2}\,m_K^2\,\bar I_\eta
\nonumber\\
&& \quad \;\; \;+\,\frac{12}{f^2}\,m_K^2\,(m_\pi^2+\,m_\eta^2)\,(2\,L_6-L_4)+\,\frac{8}{f^2}\,m_K^4\,(2\,L_8-L_5)\,,
\nonumber\\
&& m_\eta^2 \,= \frac{2}{3}\,B_0\,(m+2\,m_s)- \frac{1}{2\,f^2}\,m_\pi^2  \,\bar I_\pi  
- \frac{1}{6\,f^2}\,\Big\{7\,m_\eta^2-\,4\,m_K^2\Big\} \bar I_\eta +\,\frac{4}{3\,f^2}\,m_K^2  \,\bar I_K
\nonumber\\
&& \quad \;\;\;  +\, \frac{24}{f^2}\,m_\eta^2\,( 2\,m_K^2-\,m_\eta^2)\,(2\,L_6-L_4) 
+\,\frac{8}{f^2}\,m_\eta^4\, (2\,L_8-L_5) 
\nonumber\\
&& \quad \;\; \;+\,\frac{16}{5\,f^2}\,(3\,m_\pi^4-\,8\,m_K^4-\,8\,m_\eta^2\,m_K^2+\,13\,m_\eta^4)\,(3\,L_7+L_8) \,,
\nonumber\\
&& \bar I_Q =\frac{m_Q^2}{(4\,\pi)^2}\,
\log \left( \frac{m_Q^2}{\mu^2}\right)\,,
\label{meson-masses-q4}
\end{eqnarray} 
which involve particular combinations, $2\,L_6-L_4, 2\,L_8-L_5, 3\,L_7 + L_8$ of the LEC and 
the renormalized mesonic tadpole integral $\bar I_Q$. Our expressions differ from the traditional results of Gasser and Leutwyler \cite{Gasser:1984gg}. An expansion of (\ref{meson-masses-q4}) in  powers of the quark masses, however, recovers the traditional result upon neglect of terms cubic (or higher) in the quark masses. To this extent we suggest a particular summation scheme of the $\chi$PT approach. The rationale behind this is the use of on-shell masses inside loop contributions, however, in a manner that keeps the renormalization-scale invariance of the approach and the chiral Ward identities. Indeed, in (\ref{meson-masses-q4}) the $\mu$ dependence from the tadpole integral $\bar I_Q$ is cancelled identically by a  corresponding $\mu$ dependence in the LEC.

For the chiral Lagrangian the various contributions to the baryon self energies are well documented in the literature \cite{Semke:2005sn,Semke:2011ez,Semke:2012gs,Lutz:2012mq,Lutz:2014oxa,Lutz:2018cqo}. 
For instance tree-level contributions are detailed in Appendix A of \cite{Semke:2011ez}. There are terms linear and quadratic in the quark masses, proportional to $b_{\cdots}, c_{\cdots}$ and $d_{\cdots}, e_{\cdots}$ for the baryon octet and decuplet states respectively. Further contributions from wave-function terms proportional to $\zeta_{\cdots}, \xi_{\cdots
 }$ are linear in the quark masses. 
A first complete collection for the form of the bubble-loop diagrams can be found in \cite{Semke:2005sn,Semke:2011ez}. Most useful for the following discussion is Eq.  (30) and Eq. (31) of \cite{Lutz:2018cqo}, which specify the tadpole-type and bubble-type loop contributions. The pertinent finite box effects are documented in \cite{Lutz:2014oxa}.
The generic form of the baryon self energy of type $B$ takes the form
\begin{eqnarray}
&&  M_B - M^{(0)}_B - \bar \Sigma_B^{(2-\rm ct)}  - \bar \Sigma^{(4-{\rm ct})}_{B } 
- \bar \Sigma^{\rm tadpole}_{B }  -  \bar \Sigma^{{\rm bubble} }_B / Z_B = 0\,,
\nonumber\\
&&  Z_B = \Big(1 + \frac{\partial }{\partial M_B}\,\bar \Sigma^{\rm bubble}_B \Big)/ \Big( 1- \frac{\partial }{\partial M_B}\,\Sigma^{(4-{\rm ct})}_B\Big)\,,
\label{gap-equation-B}
\end{eqnarray}
where we denote with $\bar \Sigma_B^{(2-\rm ct)}$ and $ \bar \Sigma^{(4-{\rm ct})}_{B }$ tree-level contributions 
proportional to either a quark mass or a product of two quark masses. With $M_B^{(0)}$ we recall the baryon mass in the chiral limit with $m_u=m_d = m_s =0$. 

The  meson tadpole terms 
$\bar \Sigma^{\rm tadpole}_{B } $ are proportional to the two-body LEC $b_{\cdots}, \bar g^{(S)}_{\cdots}, 
\bar g^{(V)}_{\cdots}$ and   $d_{\cdots}, \bar h^{(S)}_{\cdots}, 
\bar h^{(V)}_{\cdots}$ for the baryon octet and decuplet states respectively, where we 
use renormalized LEC  as recalled in the Appendix for the readers' convenience. 
In the the notations of \cite{Semke:2011ez} such terms are
\begin{eqnarray}
&& \bar \Sigma^{\rm tadpole}_{B} = \frac{1}{(2\,f)^2}\sum_{Q\in [8]} \Big(  G^{(\chi )}_{BQ} 
- m_Q^2\,G^{(S)}_{BQ} - \frac 14 \, m_Q^2\,M^{(0)}_B \,G^{(V)}_{BQ}\Big)\, \bar I_Q \,,
\label{def-tadpole} 
\end{eqnarray}
where the symmetry breaking term $G^{(\chi )}_{BQ} $ is proportional to the product of a quark mass and the LEC $b_{\cdots}, d_{\cdots}$. While the Clebsch $G^{(S)}_{BQ} $ probe the scalar-type LEC  $\bar g^{(S)}_{\cdots}, \bar h^{(S)}_{\cdots}$, the term $G^{(V)}_{BQ} $
the vector-type LEC $\bar g^{(V)}_{\cdots}, \bar h^{(V)}_{\cdots}$. 
The renormalization-scale dependence from the tadpole terms can be absorbed into $\Sigma^{(4-{\rm ct})}_{B } $ upon the request of a particular $\mu$ dependence 
of the LEC with 
\begin{eqnarray}
&& \mu^2\,\frac{d }{d \,\mu^2} \,c_i = -\frac{1}{4}\,\frac{ \Gamma_{c_i}}{(4\,\pi\,f )^2}\,, \qquad \qquad 
 \mu^2 \,\frac{d}{d\,\mu^2}\,e_i = -\frac{1}{4}\,\frac{\Gamma_{e_i}}{(4\,\pi\,f )^2} \,, 
 \label{res-running}
\end{eqnarray}
where we recall the specific form of the $\Gamma_{c_i}$ and $\Gamma_{e_i}$ in the Appendix.
In analogy to our rewrite of the meson mass equation (\ref{meson-masses-q4}) we recast the quark mass factors, $m^2, m_s^2$ and $m\,m_s$, in $\bar \Sigma^{(4-ct)}$ into sums over $m_Q^4$ in such a manner that we obtain strictly scale-independent expressions. The details are specified in Tab. 1 and Tab. 2 of \cite{Lutz:2018cqo}.

We turn to the bubble-loop contributions. From 
\cite{Semke:2005sn,Semke:2011ez,Semke:2012gs,Lutz:2012mq,Lutz:2014oxa} it is straightforward to identify the expressions for the baryon octet and decuplet states in the notations of 
\cite{Semke:2005sn,Semke:2011ez}  with
\allowdisplaybreaks[1]\begin{eqnarray}
&&\bar \Sigma^{\rm bubble}_{B \in [8]} = \sum_{Q\in [8], R\in [8]}
\left(\frac{G_{QR}^{(B)}}{2\,f} \right)^2  \Bigg\{ \frac{M^2_R-M^2_B}{2\,M_B}\, \Big( \bar I_Q - m_Q^2\,\bar I_R/M_R^2 \Big)
\nonumber\\
&& \qquad \qquad \qquad  
- \,\frac{(M_B+M_R)^2}{E_R+M_R}\, p^2_{QR}\,\bar I_{QR} 
\Bigg\}
\nonumber \\
&& \qquad  \;\,\,\,+\sum_{Q\in [8], R\in [10]}
\left(\frac{G_{QR}^{(B)}}{2\,f} \right)^2 \, \Bigg\{
 \frac{(M_R+M_B)^2}{12\,M_B\,M^2_R}\,\Big(M^2_R-M^2_B\Big)\,\Big( \bar I_Q - m_Q^2\,\bar I_R/M_R^2 \Big)
\nonumber\\
&& \qquad \qquad \qquad  
 -\, \frac{2}{3}\,\frac{M_B^2}{M_R^2}\,\big(E_R+M_R\big)\,p_{QR}^{\,2}\,
\bar I_{QR} + \frac{4}{3}\,\alpha^{(B)}_{QR}
   \Bigg\}\,,
\label{result-loop-8} \\ \nonumber\\
&& p_{Q R}^2 =
\frac{M_B^2}{4}-\frac{M_R^2+m_Q^2}{2}+\frac{(M_R^2-m_Q^2)^2}{4\,M_B^2} \,,\qquad \qquad 
E_R^2=M_R^2+p_{QR}^2 \,,
\label{def-pQR}\\ \nonumber\\
&&\bar \Sigma^{\rm bubble}_{B\in [10]} = \sum_{Q\in [8], R\in [8]}
\left(\frac{G_{QR}^{(B)}}{2\,f} \right)^2  \Bigg\{ 
 \frac{(M_R+M_B)^2}{24\,M^3_B}\,\Big(M^2_R-M^2_B\Big)\,\Big( \bar I_Q - m_Q^2\,\bar I_R/M_R^2 \Big)
\nonumber\\
&& \qquad \qquad \qquad 
-\,\frac{1}{3}\,\big( E_R +M_R\big)\,p_{QR}^{\,2}\,
\bar I_{QR}+ \frac{2}{3}\,\alpha^{(B)}_{QR}
\Bigg\}
\nonumber\\
&& \qquad \;\,\,\,+\sum_{Q\in [8], R\in [10]}
\left(\frac{G_{QR}^{(B)}}{2\,f} \right)^2 \, \Bigg\{
- \frac{ M_B^2 + M_R^2}{18\,M^2_B\,M_R}\,\Big(M^2_R-M^2_B\Big)\,\Big( \bar I_Q - m_Q^2\,\bar I_R/M_R^2 \Big)
\nonumber\\
&& \qquad \qquad \qquad 
+\,\frac{M_R^4+M_B^4  + 12\,M_R^2\,M_B^2 }{36\,M^3_B\,M_R^2}\,\Big(M^2_R-M^2_B\Big)\,\Big( \bar I_Q - m_Q^2\,\bar I_R/M_R^2 \Big)
\nonumber\\
&& \qquad \qquad \qquad 
-\,\frac{(M_B+M_R)^2}{9\,M_R^2}\,\frac{2\,E_R\,(E_R-M_R)+5\,M_R^2}{E_R+M_R}\,
p_{QR}^{\,2}\,\bar I_{QR} 
\Bigg\}\,,
\label{result-loop-10}
\end{eqnarray}
where the sums in (\ref{result-loop-8}, \ref{result-loop-10}) extend over the intermediate Goldstone bosons and  the baryon
octet and decuplet states. The coupling constants $G_{QR}^{(B)}$  are determined by the axial-vector 
coupling constants $F,D,C,H$ of the baryon states in (\ref{def-L1}). The renormalized scalar bubble loop integral, $\bar I_{Q R}(p^2)$ with $p^2_{QR}$ at $p^2=M_B^2$ from (\ref{def-pQR}),  takes the form
\begin{eqnarray}
&& \bar I_{Q R}=\frac{1}{16\,\pi^2}
\left\{ \gamma^R_{B} + \frac{1}{2}\, \left(\frac{M_R^2-m_Q^2}{M_B^2}-1
\right)
\,\log \left( \frac{m_Q^2}{M_R^2}\right)
\right.
\nonumber\\
&& +\left.
\frac{p_{Q R}}{M_B}\,
\left( \log \left(1-\frac{M_B^2-2\,p_{Q R}\,M_B}{m_Q^2+M_R^2} \right)
-\log \left(1-\frac{M_B^2+2\,p_{Q R}\,M_B}{m_Q^2+M_R^2} \right)\right)
\right\}\;,
\nonumber\\
&& {\rm with } \qquad  \gamma^R_{B} = -  \lim_{m, m_s\to 0}\,\frac{M_R^2-M_B^2}{M_B^2}\,\log \left|\frac{M_R^2-M_B^2}{M_R^2}\right| \,,
\label{def-master-loop}
\end{eqnarray}
where the subtraction $\gamma^R_B$ is critical in the chiral domain with $m_Q/\Delta \ll 1$. Then it holds $\bar I_{QR} \sim m_Q^2/\Delta$ and therefore the terms (\ref{result-loop-8}, \ref{result-loop-10}) do not affect the baryon masses in the chiral limit.

The additional subtraction term $\alpha^{(B)}_{QR}$ in (\ref{result-loop-8}) and  
(\ref{result-loop-10})  has various implications. It prevents a renormalization of the LEC $b_0, b_D, b_F $ and $d_0, d_D$ in (\ref{def-Q2-terms}), but also leads to wave-function renormalization factors for all baryon states of one in the chiral limit. We detail their  specific form 
\begin{eqnarray}
&& \alpha^{(B\in\,[8])}_{QR}\, = \frac{\alpha_1\,\Delta^2}{(4\,\pi)^2} \Bigg\{ 
- \Big( M_B - M \Big)\, \Big( \frac{\Delta\,\partial}{\partial\,\Delta} -\frac{\Delta\,\partial}{\partial\,M} 
+ \frac{M+ \Delta}{M} \Big)
\nonumber\\
&& \qquad  +\, \Big( M_R - M -\Delta  \Big)\, \Big( \frac{\Delta\,\partial}{\partial\,\Delta} 
+ 1 \Big) \, \Bigg\}\,\gamma_1 
 +  \frac{\Delta\, m_Q^2}{(4\,\pi)^2}\,\alpha_1\,\gamma_2 \,, 
\nonumber\\
&& \alpha^{(B\in[10])}_{QR} = \frac{\beta_1\,\Delta^2}{(4\,\pi)^2} \Bigg\{ 
+\Big( M_B - M - \Delta\Big)\, \Big( \frac{\Delta\,\partial}{\partial\,\Delta} + 1\Big)
\nonumber\\
&& \qquad  -\,\Big( M_R - M \Big)\, \Big( \frac{\Delta\,\partial}{\partial\,\Delta} - \frac{\Delta\,\partial}{\partial\,M} 
+ \frac{M+ \Delta}{M}\Big)
\Bigg\}\,\delta_1 
 + \frac{\Delta\,m_Q^2}{(4\,\pi)^2}\,\beta_1\,\delta_2 \,, 
\label{def-alphaBR}
\end{eqnarray}
in terms of dimensionless parameters $\alpha_n, \beta_n$ and $\gamma_n, \delta_n$ depending  on the ratio $\Delta/M$ only. They are derived in Appendix A and Appendix B of \cite{Lutz:2018cqo}. While the $\alpha_n $ and $\beta_n $ are rational functions in $\Delta/M$ 
properly normalized to one in the limit $\Delta \to 0$, the  $\gamma_n$ and $\delta_n $
have a more complicated form involving terms proportional to $\log (\Delta/M)$.  As was emphasized in \cite{Lutz:2018cqo}  an expansion of such coefficients in powers of $\Delta/M$ is futile if truncated at low orders. At realistic values for $\Delta$ and $M$ the coefficients $\alpha_n$ and $\beta_n$ depart strongly from their limit value one, with even flipped signs for some cases. 

\clearpage

\section{Baryon masses on Lattice QCD ensembles }

We use first the Lattice QCD results for meson and baryon masses on various CLS ensembles as made available by the Regensburg group in \cite{RQCD:2022xux}. For given set of ensembles at fixed value of $\beta $ we determine an associated lattice scale, $a$, by enforcing that in the 
continuum limit at physical quark masses and infinite box size the isospin-averaged baryon octet and decuplet masses are reproduced. 

In addition we consider discretization effects by permitting LEC to depend on the lattice scale $a$. 
While it cannot be proven that such an ansatz grasps all discretization effects with sufficient accuracy, we would 
argue that a QCD action which is not consistent with this assumption, is at least unfortunate and would ask for modifications. From a pragmatic point of view, in the absence of detailed knowledge on the properties of the used action, we deem our strategy to be sufficiently well motivated. For a given mass parameter $M_{\rm hadron}$ we form the particular combination 
\begin{eqnarray}
&&a\,M_{\rm hadron}\,(1 - a\,\bar m\,b_a  )  \,,\qquad  \qquad \qquad 
b_a =  0.31583(5) + {\rm \mathcal O}{ (1/\beta ) } \,,
\label{def-combinations}
\end{eqnarray}
with values for $a \,\bar m$ listed in \cite{RQCD:2022xux} for the various ensembles. According to  \cite{RQCD:2022xux} the considered combination 
(\ref{def-combinations}) receives further corrections from discretization effects at least quadratic in the lattice scale $a$ only.  The parameter $b_a $ is so far only known at the one-loop level \cite{RQCD:2022xux}. Therefore, we consider the value for $b_a$ as a free parameter which we adjust to the data set, where we search in the restricted range $-1 < b_a <1$, that includes the somewhat smaller range suggested in \cite{RQCD:2022xux}.  Note that $b_a$ can only be determiend, if lattice data 
for given $\beta $ value come with various values of $\bar m$. 

In turn, we can model discretization effects by using LEC that have a quadratic lattice scale dependence. We do so only for the leading order LEC, such as 
\begin{eqnarray}
&& m  \;\;\,\,\to m\,\big( 1+ a^2\,\gamma_{ud} \big) \,,\quad \quad m_s \to m_s\,\big( 1+ a^2\,\gamma_s \big) \,,
\nonumber\\
&& M_{[8]} \,\to M_{[8]} \,+ a^2\,\gamma_{M_8} \,,   \qquad \;
 b_0 \to b_0 + a^2\,\gamma_{b_0} \,,\qquad \! \! \!
 b_D \to b_F + a^2\,\gamma_{b_D} \,,\qquad \! \!
  b_F \to b_F + a^2\,\gamma_{b_F} \,,
  \nonumber\\
&& M_{[10]} \to M_{[10]} + a^2\,\gamma_{M_{10}} \,,
 \qquad \!  \!d_0 \to d_0 + a^2\,\gamma_{d_0} \,,\qquad \! \! \!
 d_D \to d_D + a^2\,\gamma_{d_D} \,,
 \label{def-gamma}
\end{eqnarray}
where the set of parameters $\gamma_{\cdots}$ depends on the QCD action used. We note that such effects can be generated by following 
the strategy developed in \cite{Sharpe:1998xm,Rupak:2002sm,Aoki:2005mb,Aoki:2006ab} for a Wilson quark action. Since by now QCD actions are order $a$ improved, it is justified to drop linear spurion field structures that are proportional to $a$. The particular form for the quark masses shown in (\ref{def-gamma}) is implied by Eqs. (24)-(26) in  \cite{Aoki:2005mb}, where we translated the $a^2$ structures $C_0$ and $D_0$ into our notation $\gamma_{ud}$ and $\gamma_s$. As a consequence of that particular form the discretization parameters, $\gamma_{ud}$ and $\gamma_s$, cannot be  determined in our current work. 


Our computation uses the pion and kaon masses as given in lattice units to determine the quark masses, $m, m_s$ (including their possible $a^2$ dependence) together with the eta mass by solving the finite-box variant of set of nonlinear equations (\ref{meson-masses-q4}). This involves two parameter combinations, $2\,L_6-L_4$ and $2\,L_8 -L_5$, where we recall that $3\,L_7 +L_8$ is determined by the request that the empirical eta mass is reproduced at physical values of the quark masses. With this, we can compute the baryon masses in terms of the finite-box variant 
of the set of nonlinear equations (\ref{gap-equation-B}).  This involves a series of LEC 
as discussed in Section II. 

\begin{table}[t]
\setlength{\tabcolsep}{4.5mm}
\renewcommand{\arraystretch}{1.3}
\begin{center}
\begin{tabular}{l|rr}
                                 & Fit (to CLS \cite{RQCD:2022xux})    & Fit (to world \cite{Guo:2019nyp})          \\   \hline

$10^3\,(2\,L_6 -L_4 )\, $        &  0.0411(3)         &  0.0401$^{(24)}_{(01)}$         \\
$10^3\,(2\,L_8 -L_5 )\, $        &  0.0826(12)         &  0.1049$^{(43)}_{(43)}$        \\
$10^3\,(L_8 + 3\,L_7)\, $        & -0.4768(4)         & -0.4818$^{(09)}_{(13)}$        \\ \hline 
$m_s/m$                          &  26.15(1)          &   26.02$^{(02)}_{(02)}$         \\

\end{tabular}
\caption{LEC from a fit to the baryon octet and decuplet masses. The values in the second column show the result of our current fit to the results of the Regensburg on their baryon masses. In the third column we recall our previous values \cite{Guo:2019nyp},  where 
results of the  PACS-CS, LHPC,  HSC, NPLQCD, QCDSF-UKQCD and ETMC groups on the baryon masses were used only. All parameters  $L_i$ are scale dependent 
given at $\mu =$ 770 MeV together with $f =92.4$ MeV (see (\ref{meson-masses-q4})). }
\label{tab:FitParametersA}
\end{center}
\end{table}

In Tab. \ref{tab:FitParametersA}  we present the implications of our global fit to the baryon masses on the CLS ensembles \cite{RQCD:2022xux}.  While our quark-mass ratio $m_s/m$ suffers from a rather small uncertainty only, our value is not compatible with the current FLAG report value 27.42(12) \cite{Aoki:2021kgd}. On the other hand, we find values for the Gasser and Leutwyler LEC that are quite compatible with a corresponding fit to the previous world Lattice data on the baryon masses  \cite{PACS-CS:2008bkb,HadronSpectrum:2008xlg,Walker-Loud:2008rui,Beane:2011pc,Bietenholz:2011qq,Alexandrou:2013joa}, despite the fact that no attention was paid to possible discretization effects. Our error estimate of the LEC considers the 1-sigma statistical uncertainties of our fit, but also  some systematic uncertainties as is implied  by  the limited accuracy of our chiral extrapolation framework. While the latter will reliably only be known after a full computation of the chiral extrapolation framework at the two-loop level, at this stage we assume a universal error that is estimated by the request that our Lattice data description reaches a $\chi^2/{\rm dof} \simeq 1$. This leads to an uncertainty of about 12-14 MeV, which we vary slightly around its optimal value. The total uncertainty is typically dominated by the latter variation rather than the 1-sigma statistical uncertainty, simply because the number of  fitted  data points, about 400, is quite large. 

We consider here Lattice QCD data points with pion and kaon masses equal to or smaller than the empirical eta mass only. This is the most conservative assumption we can do without jeopardizing our three-flavor extrapolation framework. A flavor dependent data selection,  where ensembles with pion or kaon masses significantly smaller than the empirical eta mass are rejected, is hard to justify, at least if the target is to extrapolate the framework towards the physical point.  

\begin{table}[t]
\setlength{\tabcolsep}{4.5mm}
\renewcommand{\arraystretch}{1.2}
\begin{center}
\begin{tabular}{l|c|c||l|c|c} 

                                                    &  Fit    &  CLS \cite{RQCD:2022xux} &   &  Fit   &  CLS \cite{RQCD:2022xux} \\ \hline 
$a^{\beta =3.34}_{\rm CLS}\,   \hfill \mathrm{[fm]}$       &  0.09337(22) & 0.09757(56)   & $a^{\beta =3.55}_{\rm CLS}\,   \hfill \mathrm{[fm]}$       &  0.06314(12)
& 0.06379(37)  \\   
$a^{\beta =3.40}_{\rm CLS}\,   \hfill \mathrm{[fm]}$       &  0.08251(7) & 0.08524(49)   &  $a^{\beta =3.70}_{\rm CLS}\,   \hfill \mathrm{[fm]}$       &  0.05003(16) & 0.04934(28)  \\   
$a^{\beta =3.46}_{\rm CLS}\,   \hfill \mathrm{[fm]}$       &  0.07478(8) & 0.07545(44)   & $a^{\beta =3.85}_{\rm CLS}\,   \hfill \mathrm{[fm]}$       &  0.03845(11) & 0.03873(22) \\ 
                                                      
\hline 
                                                   
$\gamma_{M_8}  \hfill   \mathrm{[GeV^3]}$     & -0.1322(10)  &   &  
$\gamma_{M_{10}}  \hfill  \mathrm{[GeV^3]}$   & -0.0776(4)  &    \\

$\gamma_{b_0}   \hfill  \mathrm{[GeV]}$       &  0.0619(8)  &   &  
$\gamma_{d_0}   \hfill \mathrm{[GeV]}$        & -0.0115(9)  &     \\
                                                      
$\gamma_{b_D}  \hfill  \mathrm{[GeV]}$        & -0.1512(9)  &   &  
$\gamma_{d_D}   \hfill \mathrm{[GeV]}$        &  0.0206(9)  &    \\

$\gamma_{b_F}   \hfill  \mathrm{[GeV]}$       & -0.0071(4)  &   & 
$b_a$                                         & 0.6305(8)   &    \\   \hline 


$M_{[8]}  \hfill   \mathrm{[GeV]}$            &  0.8043(9)   &  0.809$^{(71)}_{(53)}$ &
$M_{[10]}  \hfill  \mathrm{[GeV]}$            &  1.1152(1)   &  1.147$^{(74)}_{(91)}$ \\
 
$b_0   \hfill  \mathrm{[GeV^{-1}]}$           & -0.8144(9)   &  -0.706$^{(56)}_{(69)}$ &  
$d_0   \hfill \mathrm{[GeV^{1}]}$             & -0.4347(14)   &  -0.84$^{(44)}_{(34)}$ \\
                                                      
$b_D  \hfill  \mathrm{[GeV^{-1}]}$            &  0.1235(2)   &  0.083$^{(33)}_{(35)}$ &  
$d_D  \hfill \mathrm{[GeV^{-1}]}$             & -0.5169(13)   &  -0.50$^{(18)}_{(96)}$  \\

$b_F  \hfill  \mathrm{[GeV^{-1}]}$            & -0.2820(3)   & -0.384$^{(28)}_{(44)}$  & 
                                              &    &    \\   
                                                      
\end{tabular}
\caption{Our determination of the lattice scales for the CLS ensembles together with the discretization parameters introduced in (\ref{def-combinations}) and  (\ref{def-gamma}). We use $f = 92.4$ MeV and  $F =0.51, D = 0.72$ together with the relation (\ref{res-LargeNc}) for $C$ and $H$ in our study always.  
 }
\label{tab:lattice-scale}
\end{center}
\end{table}

More details on the fit itinerary are collected in Tab. \ref{tab:lattice-scale}, where we present our result for the lattice scale parameters together with the discretization parameters $\gamma_{\cdots }$ that drive the size of discretization effects. It is worth mentioning that the lattice scales we obtained in our quite unconventional scale-setting program agree  better and better with the values obtained in  \cite{RQCD:2022xux}  as the $\beta$ value rises, implying  a  more and more continuum-like system. This is to be expected. We find interesting the size of the parameter $b_a \simeq 0.63$ in Tab. \ref{tab:lattice-scale}, which is almost twice as large as its one-loop size. That is a strong hint that effects linear in the lattice scale are important for a quantitative reproduction 
of the data set. For the readers convenience the table includes also the set of LEC, for which the discretization effects are assumed. Upon an inspection of 
(\ref{def-gamma}) the impact of the $\gamma$s on the LEC with smaller than 5\% is moderate in almost all  cases on our coarsest lattice. The largest relative change we see in $b_D$, which, however, is already quite small. 
A quantitative comparison with the size of discretization effects as implemented in \cite{RQCD:2022xux} is quite difficult for us, since on the one hand 
the form imposed by the Regensburg group is not compatible with our ansatz (\ref{def-gamma}) and 
also no numerical values for their discretization parameters substantiating their approach are documented. 

Nevertheless, a comparison of our LEC in the Tab. \ref{tab:lattice-scale} with corresponding values from \cite{RQCD:2022xux}, has an interesting message. While the chiral limit values of the octet and decuplet masses are remarkably compatible, this is somewhat less the case for the subleading LEC, where some moderate tension even given their uncertainty estimates persists. Here we compare with the setup 'BChPT FV SC$\small \infty $' of  \cite{RQCD:2022xux}, that includes the octet and decuplet baryons, as we do in our framework also. It should be mentioned that the extrapolation scheme used in  \cite{RQCD:2022xux} is largely incompatible 
with our scheme, as was discussed in depth in our previous work \cite{Lutz:2018cqo}. 
We suspect that the differences in the extrapolation schemes can be compensated for in part by distinct treatments of 
discretization effects. It would be important to consolidate the form of the used extrapolation scheme. 
As we will show further  below this will have important consequences for some quantities at the physical point.

\begin{figure}[t]
\center{
\includegraphics[keepaspectratio,width=0.8\textwidth]{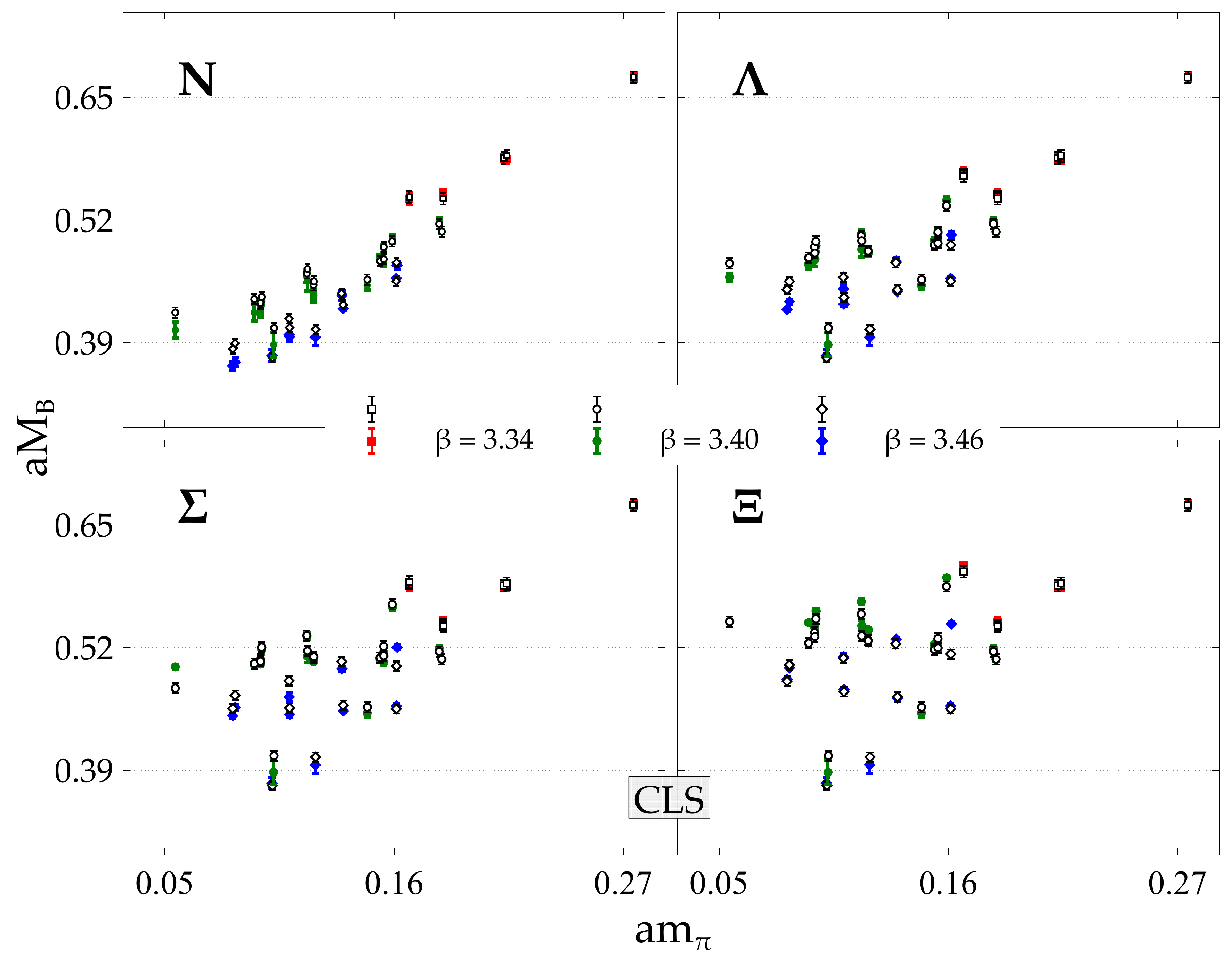} 
\includegraphics[keepaspectratio,width=0.8\textwidth]{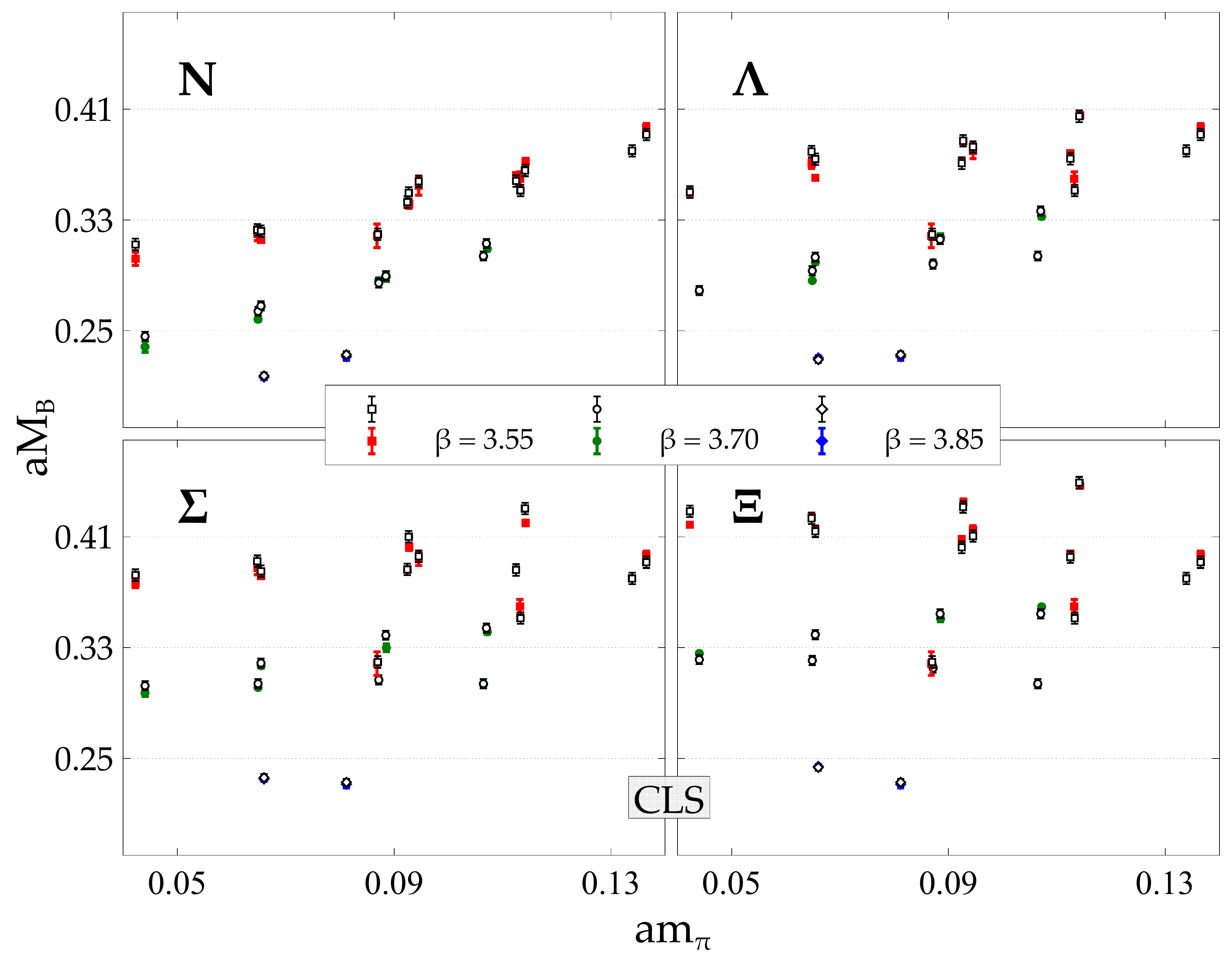} }
\vskip-0.2cm
\caption{\label{fig:13} Baryon octet masses on CLS ensembles as a function of the pion mass (in lattice units).  }
\end{figure}
 
\begin{figure}[t]
\center{
\includegraphics[keepaspectratio,width=0.8\textwidth]{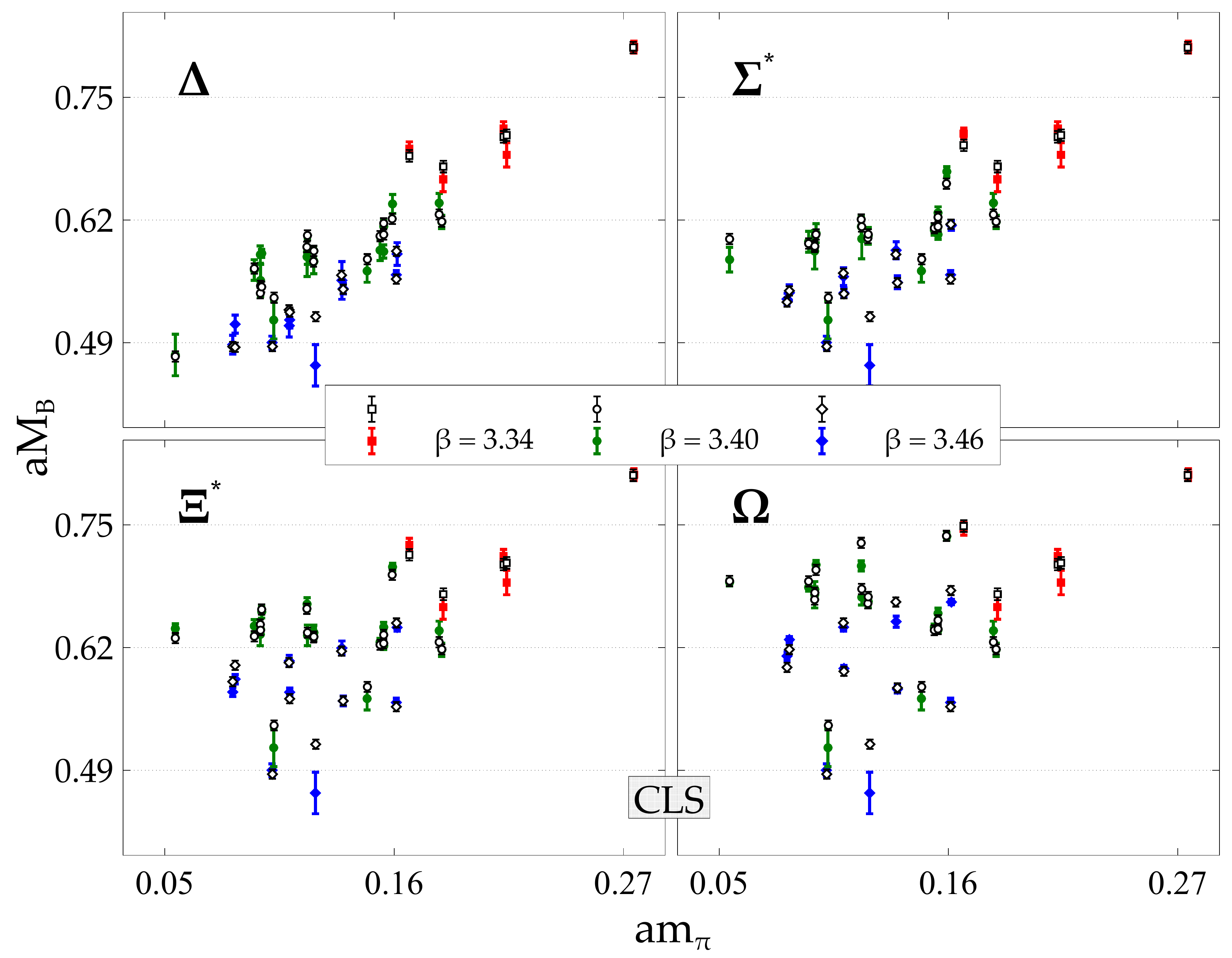} 
\includegraphics[keepaspectratio,width=0.8\textwidth]{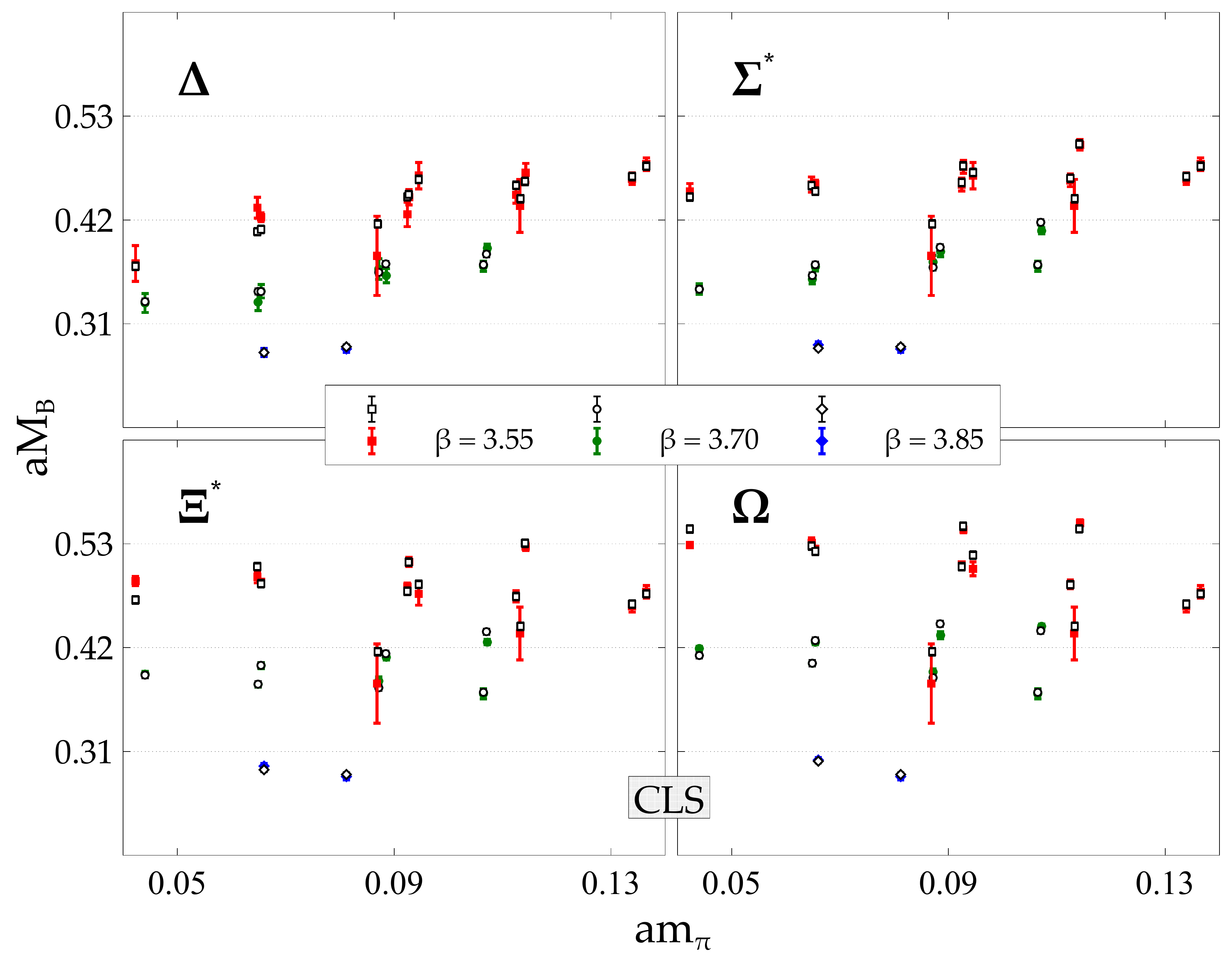} }
\vskip-0.2cm
\caption{\label{fig:24} Baryon decuplet masses on CLS ensembles as a function of the pion mass (in lattice units).  }
\end{figure}

\clearpage

We support the findings of the Regensburg group, that the data set on the baryon masses does not 
constrain $f$ and $F, D$ in a significant manner. A free fit would suggest unreasonably small values for $F$ and $D$ or large values for $f$. Therefore we perform our fit at fixed values thereof. A change in such parameters can be compensated for in part by 
our extended set of LEC.

To resolve that issue a combined fit that considers the baryon masses together with the axial form factors of the baryons would useful. While we prepared a suitable  chiral extrapolation framework for the form factors already in \cite{Lutz:2020dfi,Sauerwein:2021jxb}, there is not yet a sufficient Lattice QCD data base for this form factors available.

We illustrate the quality of our Lattice QCD data reproduction in Fig. \ref{fig:13} and Fig. \ref{fig:24} for the baryon octet and decuplet states, as a function of the pion mass. All masses  are given in lattice units. The lattice data are shown in colored symbols always, where a given color corresponds to ensembles at a fixed $\beta$ value. The chiral EFT results are in white symbols, where the shown error size is the 13 MeV estimate as discussed above.

In overall an excellent reproduction of the data on all ensembles is shown, with some tension typically on the ensembles with the smallest pion masses. With the nucleon and lambda masses on the D450 and D451 ensembles we would identify outlier points with a significant contribution to our chisquare function. Indeed, in our final fits those points were excluded. 
Here we should state that we excluded the baryon masses on the ensembles D150 and E250 from our chisquare. Both correspond to a pion mass of about 130 MeV, smaller than the empirical value,  where the isobar state is characterized most likely by more than one finite-box energy level. In this case, our request to use on-shell masses for the baryons inside loop contributions 
turns ambiguous, and the framework would need an extension. The situation may appear strange at first, since we argue that our scheme is applicable for small box sizes and in the limit of infinite box sizes. This is so since at intermediate size boxes the isobar requires a multi-channel setup, which is beyond our current framework. In the infinite volume limit, instead, the quasi-particle approximation for the isobar propagator in the loop contributions is well applicable. Note that this issue was already discussed along our line for the case of the $\rho$ meson on Lattice QCD ensembles in \cite{Guo:2018zvl}. It is  should be clear from our argument that it is quite a challenge to extrapolate from a finite box and small pion-mass ensemble to the 
infinite volume case, at least for systems where loop contributions involving an isobar propagator play an important role (see also \cite{Bulava:2021fre}). Indeed, this was already shown for the case of the axial-vector form factor of the nucleon, for which the volume dependence of the isobar mode was shown to strongly influence the volume dependence of the form factor \cite{Lutz:2020dfi}. 

\begin{table}[t]
\setlength{\tabcolsep}{4.5mm}
\renewcommand{\arraystretch}{1.2}
\begin{center}
\begin{tabular}{lr|lr|lr}
 
$g^{(S)}_0\,\hfill\mathrm{[GeV^{-1}]}$       &  -7.1454(14) &  $ g^{(V)}_0\,\hfill\mathrm{[GeV^{-2}]}$   &  1.5983(7)  &  $c_0\,\hfill\mathrm{[GeV^{-3}]}$      &  0.3884(8)       \\
$ g^{(S)}_1\,\hfill\mathrm{[GeV^{-1}]}$       &  -0.6454(1) & $ g^{(V)}_1\,\hfill\mathrm{[GeV^{-2}]}$    & -4.1566(12)  &  $\zeta_0\,\hfill\mathrm{[GeV^{-2}]}$  &   0.3823(14)   \\
$ g^{(S)}_D\,\hfill\mathrm{[GeV^{-1}]}$       & -0.8087(14) &  $ g^{(V)}_D\,\hfill\mathrm{[GeV^{-2}]}$    &  3.2166(10)   &  $\zeta_D\,\hfill\mathrm{[GeV^{-2}]}$  &  -0.1145(5)  \\
$ g^{(S)}_F\,\hfill\mathrm{[GeV^{-1}]}$       & -2.9798(9) &  $ g^{(V)}_F\,\hfill\mathrm{[GeV^{-2}]}$    & -3.9400(3) &  $\zeta_F\,\hfill\mathrm{[GeV^{-2}]}$   & 0.0504(3) \\ 

$c_1\,\hfill\mathrm{[GeV^{-3}]}$   & 0.2201(7)    &  $c_2\,\hfill\mathrm{[GeV^{-3}]}$    &  -0.4521(3)    &
$c_3\,\hfill\mathrm{[GeV^{-3}]}$   & 0.4591(6)   \\
$c_4\,\hfill\mathrm{[GeV^{-3}]}$   & 0.3281(3)   & $c_5\,\hfill\mathrm{[GeV^{-3}]}$      &  -0.3176(3)   & 
$c_6\,\hfill\mathrm{[GeV^{-3}]}$   &  -0.2893(5) \\

\end{tabular}
\caption{LEC from a fit to the CLS baryon masses at $\mu =701$ MeV. All LEC in this table may be considered as independent.
}
\label{tab:FitParametersB}
\end{center}
\end{table}

\begin{table}[b]
\setlength{\tabcolsep}{4.5mm}
\renewcommand{\arraystretch}{1.2}
\begin{center}
\begin{tabular}{lr|lr|lr}
$   h^{(S)}_1\,\hfill\mathrm{[GeV^{-1}]}$       & -3.0444(2) &  $   h^{(V)}_1\,\hfill\mathrm{[GeV^{-2}]}$   & 3.8897(1)  &  $e_0\,\hfill\mathrm{[GeV^{-3}]}$      & -0.2560(9)   \\
$   h^{(S)}_2\,\hfill\mathrm{[GeV^{-1}]}$       &  0.     & $   h^{(V)}_2\,\hfill\mathrm{[GeV^{-2}]}$    & -0.4844(82)  &  $e_1\,\hfill\mathrm{[GeV^{-2}]}$      &  0.3806(75)   \\
$   h^{(S)}_3\,\hfill\mathrm{[GeV^{-1}]}$       & -1.9280(2) &  $   h^{(V)}_3\,\hfill\mathrm{[GeV^{-2}]}$   &  -4.8073(60)   &  $e_2\,\hfill\mathrm{[GeV^{-2}]}$      &   0.3005(83)  \\
$   h^{(S)}_4\,\hfill\mathrm{[GeV^{-1}]}$       & -1.7500(7) &  $\xi_0\,\hfill\mathrm{[GeV^{-2}]}$          & 
0.2174(11) &  $e_3\,\hfill\mathrm{[GeV^{-2}]}$       & 0.0314(1)    \\
$   h^{(S)}_5\,\hfill\mathrm{[GeV^{-3}]}$       & -4.9384(10) &  $\xi_D\,\hfill\mathrm{[GeV^{-3}]}$         &  
-0.1923(9)  & $e_4\,\hfill\mathrm{[GeV^{-2}]}$       & 0.0897(6) \\

$h^{(S)}_6\,\hfill\mathrm{[GeV^{-3}]}$   &   1.7500(7)    & & \\
\end{tabular}
\caption{LEC from a fit to the CLS baryon masses  at $\mu =701$ MeV. The 16 LEC in this table are strongly correlated by the large-$N_c$ sum rules in (\ref{res-LargeNc}). There remain only four LEC that are independent of those in Tab. \ref{tab:FitParametersB}.
}
\label{tab:FitParametersC}
\end{center}
\end{table}

We complete our list of LEC with Tab. \ref{tab:FitParametersB} and Tab. \ref{tab:FitParametersC} in the octet and decuplet sectors respectively.  All LEC are well determined in our fit and have natural size. Their values differ in part significantly from our previous results in  \cite{Lutz:2018cqo,Guo:2019nyp}, which we attribute mainly to the neglect of discretization effects in our previous studies. 

\clearpage

\begin{figure}[t]
\center{
\includegraphics[keepaspectratio,width=0.8\textwidth]{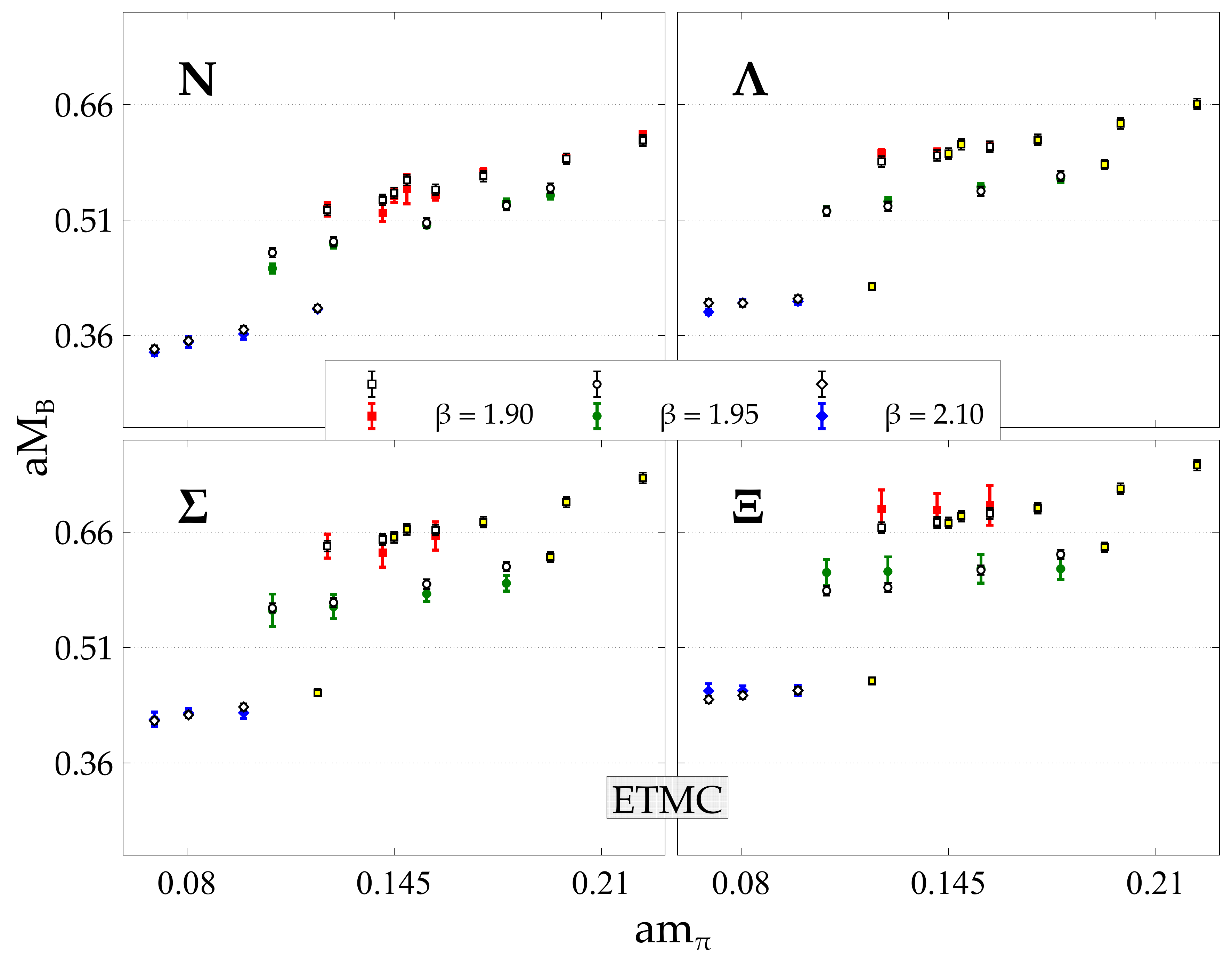} 
\includegraphics[keepaspectratio,width=0.8\textwidth]{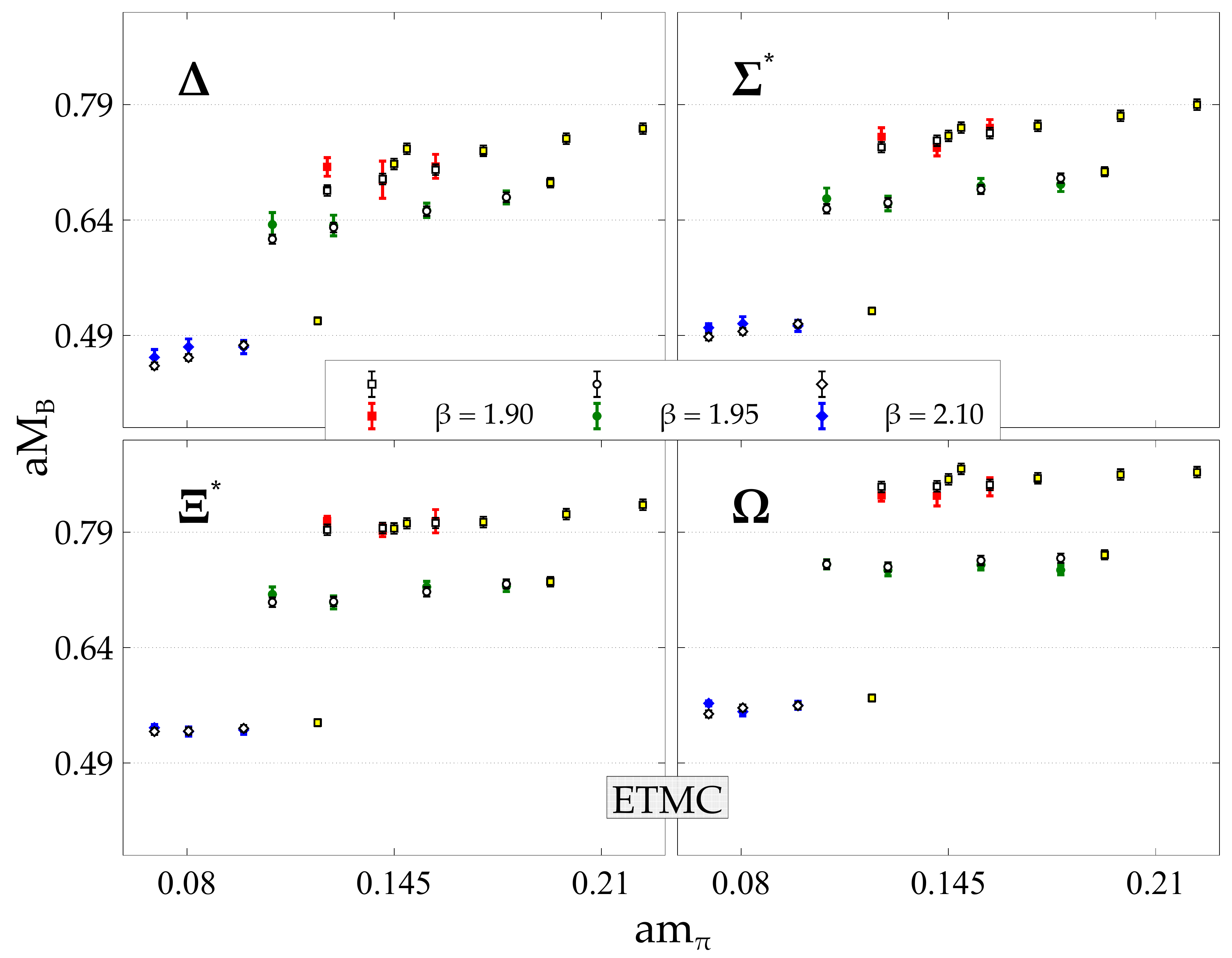} }
\vskip-0.2cm
\caption{\label{fig:78} Baryon masses on ETMC ensembles as a function of the pion mass (in lattice units). }
\end{figure}

\begin{figure}[t]
\center{
\includegraphics[keepaspectratio,width=0.8\textwidth]{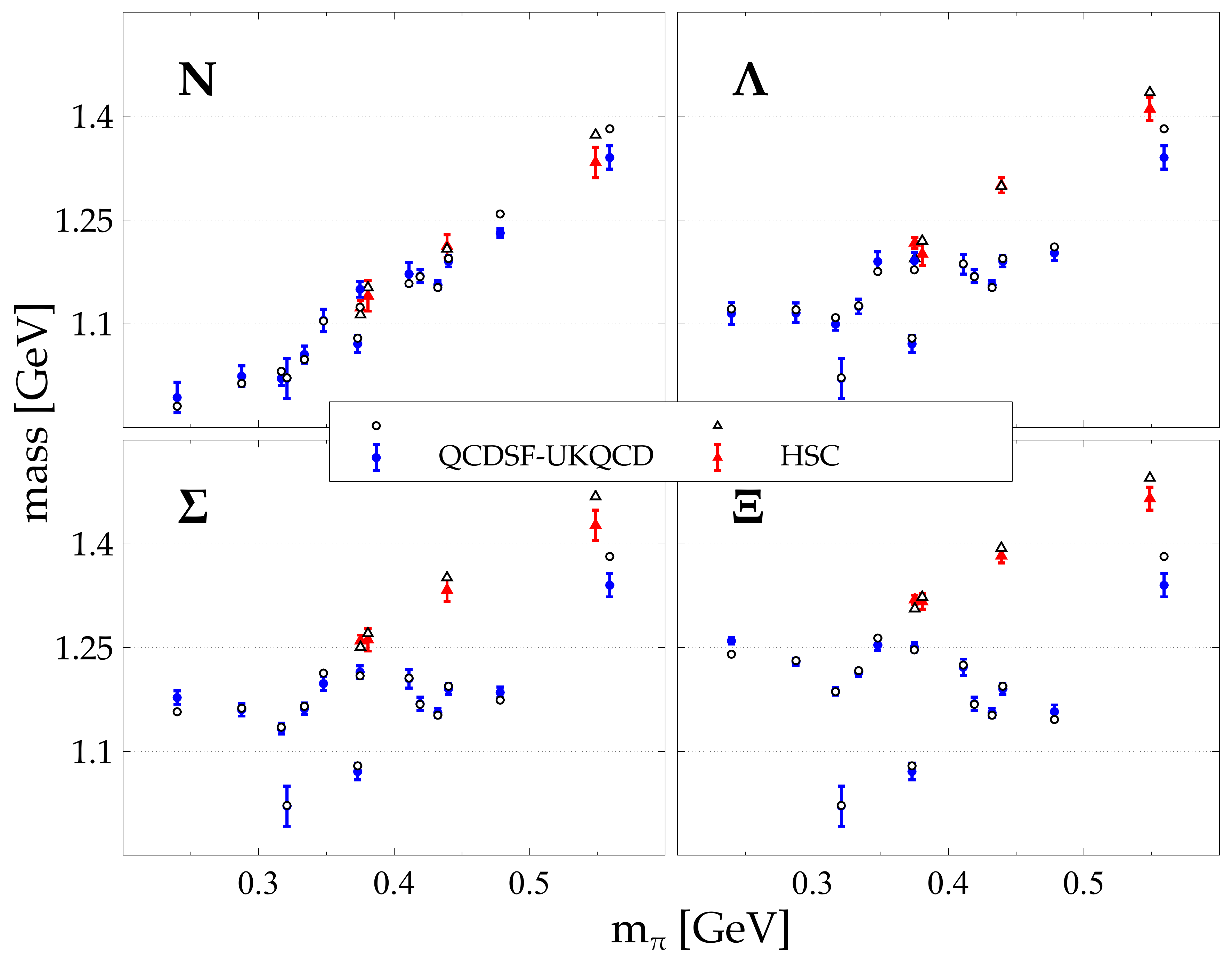} 
\includegraphics[keepaspectratio,width=0.8\textwidth]{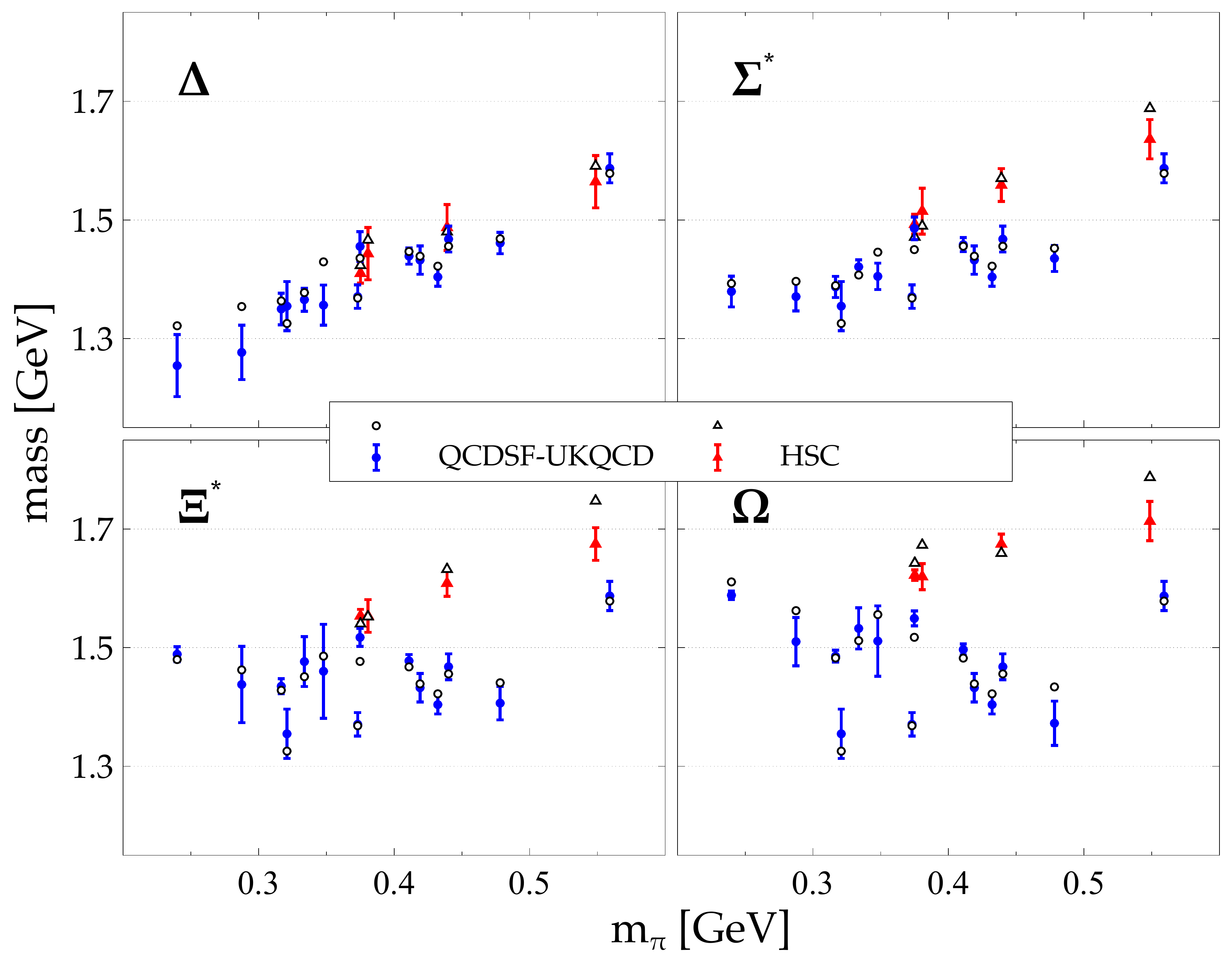} }
\vskip-0.2cm
\caption{\label{fig:56} Baryon masses on QCDSF-UKQCD and HSC ensembles as a function of the pion mass (in physical units).  }
\end{figure}

\clearpage

Since such LEC are crucial input for coupled-channel computations of meson-baryon scattering, we wish to 
further consolidate our results. In particular the reader may be worried about our claims that the LEC 
\cite{Lutz:2018cqo,Guo:2019nyp} reproduce a large class of Lattice QCD data on baryon masses. The best we can do at this stage is to 
use our new LEC of Tab. \ref{tab:FitParametersB} and Tab. \ref{tab:FitParametersC} and confront those with the previous  results by 
PACS-CS, LHPC, HSC, NPLQCD, QCDSF-UKQCD and ETMC \cite{PACS-CS:2008bkb,HadronSpectrum:2008xlg,Walker-Loud:2008rui,Beane:2011pc,Bietenholz:2011qq,Alexandrou:2013joa}.

\begin{table}[t]
\setlength{\tabcolsep}{2.3mm}
\renewcommand{\arraystretch}{1.1}
\begin{center}
\begin{tabular}{l|r|r|r||ccc} 

&  ETMC &    QCDSF & HSC &  scale & Fit  & Lattice  \\ \hline 
                                                      
\hline 
                                                   
$\gamma_{M_8}  \hfill   \mathrm{[GeV^3]}$     &   0.003(12)    &  -0.284(2)   &     -0.053(13)   &      $a^{\beta =1.90}_{\rm ETMC}\,   \hfill \mathrm{[fm]}$   &  0.105(2) & 0.0934(37)  \\
$\gamma_{b_0}   \hfill  \mathrm{[GeV]}$       &   0.201(10)    &   0.104(3)   &      0.117(25)   &      $a^{\beta =1.95}_{\rm ETMC}\,   \hfill \mathrm{[fm]}$   &  0.094(1) & 0.0820(37)  \\   
$\gamma_{b_D}  \hfill  \mathrm{[GeV]}$        &  -0.106(1)     &  -0.001(5)   &     -0.027(4)    &      $a^{\beta =2.10}_{\rm ETMC}\,   \hfill \mathrm{[fm]}$   &  0.070(1) & 0.0644(26)  \\   
$\gamma_{b_F}   \hfill  \mathrm{[GeV]}$       &  -0.112(14)     &  -0.011(4)   &      0.019(4)    &      $a^{}_{\rm QCDSF}\,   \hfill \mathrm{[fm]}$             &  0.080(1) & 0.0765(15)  \\   
$\gamma_{M_{10}} \hfill  \mathrm{[GeV^3]}$    &   0.009(13)    & -0.128(11)   &      0.117(25)   &      $a^{}_{\rm HSC}\,   \hfill \mathrm{[fm]}$               &  0.125(2) & 0.123(1)    \\ 
$\gamma_{d_0}   \hfill \mathrm{[GeV]}$        &   0.238(21)    &  0.235(8)    &      0.173(12)   &                                                              &           &             \\
$\gamma_{d_D}   \hfill \mathrm{[GeV]}$        &  -0.147(13)     & -0.109(12)   &     -0.038(2)    &                                                              &           &             \\
\end{tabular}
\caption{Our determination of the spatial  
lattice scales for the ETMC  \cite{Alexandrou:2013joa},   QCDSF-UKQCD \cite{Bietenholz:2011qq} and HSC  \cite{HadronSpectrum:2008xlg} ensembles together with the discretization parameters introduced in (\ref{def-combinations}) and  (\ref{def-gamma}). 
 }
\label{tab:gamma-scale}
\end{center}
\end{table}

Owing to the use of different QCD actions, the discretization effects in such works are distinct  to the ones found already for the CLS ensembles. Therefore, our discretization parameters $\gamma_{\cdots}$ need to be readjusted. This is possible for the data set of ETMC,  QCDSF-UKQCD and HSC, for which we show 
in Fig. \ref{fig:78}  and Fig. \ref{fig:56} results in comparison with the Lattice data as a function of the pion mass.  
The values of the discretization parameters are collected in Tab. \ref{tab:gamma-scale}.  Clearly, such a readjustment should come with an update of the scale setting. Our results for the scale parameters are shown in Tab. \ref{tab:gamma-scale} and compared with values from the lattice groups. A quite consistent pattern is observed. The ETMC masses are shown in lattice units, where again a given color reflects ensembles at fixed $\beta$ value. They are compared with white symbols implied by the chiral EFT. 
Note that for the cases where the yellow point does not come with an associated colored point, there is so far no lattice determination of the baryon mass available on that ensemble. 
The masses on the HSC and QCDSF-UKQCD ensembles are shown in physical units, where we use red color for the HSC and blue color for the QCDSF-UKQCD. The chiral EFT results are again shown in white symbols. 

After that successful test of our set of LEC we feel that it is justified to show a first physics application. We consider the pion-nucleon sigma terms of the nucleon together with its scalar strangeness content,   
\begin{eqnarray}
&& \sigma^{\rm Regensburg}_{\pi N}  = (43.9 \pm 4.7 ) \,{\rm MeV } \,,\qquad \qquad \quad \;\;\,\sigma^{\rm Regensburg}_{sN} = 16^{(58)}_{(68)} \,{\rm MeV }\,,
\nonumber\\
&&  \sigma^{\rm chiral\; EFT}_{\pi N} \; = ( 58.7 \pm 1.3 ) \,{\rm MeV } \,,\qquad \qquad \quad \; \; \,  \sigma^{\rm chiral \;EFT}_{sN} =( -316 \pm 76 )\,{\rm MeV }\,,
\label{res-compare}
\end{eqnarray}
where we confront our predictions with those of the Regensburg group, that are based on the identical Lattice QCD data set. A striking 
disagreement is seen. While our value for $\sigma_{\pi N}$ agrees well with the empirical result $\sigma_{\pi N} =58(5)$ MeV from  \cite{RuizdeElvira:2017stg,Gupta:2021ahb}, this is less so for the Regensburg group. Most striking are the quite distinct values for the strangeness content for which there is no empirical reference value available. We conclude, that it would be important to consolidate such analyses by further improvements in the Lattice QCD data base.

\section{Summary and outlook}
\label{sec:summary}

In this work we report on a first analysis of the baryon octet and decuplet masses from recent CLS ensembles based 
on a chiral extrapolation framework at next-to-next-to-next-to-leading order, where the particular summation scheme as implied by using on-shell meson and baryon masses in loop contributions was scrutinized. Values for all LEC relevant in the baryon masses at this order were determined accurately in the continuum limit. The parameter set was tested successfully against previous Lattice QCD data sets from ETMC, QCDSF-UKQCD and HSC, where the size of discretization effects was predicted. With our results the path towards coupled-channel computations for meson-baryon scattering with controlled QCD input is paved.

We pointed out that the type of chiral extrapolation method has a crucial impact on some results at the physical point, despite the fact that there exist CLS ensembles at rather small pion masses. While such differences in the extrapolation schemes can be compensated for in part by distinct treatments of discretization effects, final results may have a strong dependence on the used  scheme. For instance, our prediction for the pion-nucleon sigma term is compatible with its empirical value \cite{RuizdeElvira:2017stg}, which is not the case in the most recent findings of the Regensburg group on the CLS ensembles. Even more striking is our rather large scalar strangeness content of the nucleon, that differs from their findings in sign and also in magnitude significantly. It would be important to consolidate the form of the to-be  used chiral extrapolation scheme. 

From phenomenological studies it is expected that the isobar degree of freedom plays a significant role in the determination of the sigma terms. The challenge of lattice ensembles at rather small pion masses (even below its physical value) is the difficult-to-control finite-box effects. While in the infinite volume, a well established quasi-particle approximation for the isobar propagator inside loop contributions is applicable, for medium size boxes it is more challenging how to model such finite volume effects. As the box size is chosen, so that the  isobar is well described by a single box level the system permits better control again. Based on this observation, we find in our framework that the impact of ensembles with not-too-small pion masses is more useful for a  quantitative control of observables at the physical point with its infinite volume limit, at least for the current CLS ensembles.

\section*{Acknowledgments}

We thank John Bulava and Daniel Mohler for discussions on Lattice QCD aspects and the Regensburg group for providing us with 
tables of their numerical results on the CLS ensembles.

\section*{Appendix A}

We express the LEC  $\bar g_{\cdots}^{(S, V)}$ and  $\bar h_{\cdots}^{(S, V)}$ that characterize the tadpole-type contributions in the baryon self energy (\ref{def-tadpole}) in terms of the 
LEC introduced in the chiral Lagrangian as
\begin{eqnarray}
&& \bar g_0^{(S)} = g_0^{(S)} - \frac{4}{3}\, g^{(S)}_C \,,  \qquad \quad 
 \bar g_1^{(S)} = g_1^{(S)}  +\frac{1}{3}\,g^{(S)}_C \,,  
\nonumber\\
&& \bar g_D^{(S)} =   g_D^{(S)} + g^{(S)}_C\,,   \qquad \qquad 
 \bar g_F^{(S)} =g_F^{(S)}  - g^{(S)}_C \,,\qquad g^{(S)}_C =  \frac{2\,C^2}{3\,M}\,\alpha_5 \,,
\nonumber\\ 
&& \bar g_0^{(V)} =  g_0^{(V)}  - \frac{4}{3}\, g^{(V)}_C \,, \qquad 
\bar  g_1^{(V)} =  g_1^{(V)}   + \frac{1}{3}\,g^{(V)}_C\,,
\nonumber\\
&& \bar g_D^{(V)} =   g_D^{(V)}  + g^{(V)}_C\,,\qquad \quad 
\bar g_F^{(V)} =  g_F^{(V)} - g^{(V)}_C\,,\qquad g^{(V)}_C = \frac{C^2}{3\,M^2}\,\alpha_{6}  \,,
\nonumber\\
&&\bar  h_1^{(V)} =  h_1^{(V)}   \,, \qquad \qquad \qquad 
 \bar h_2^{(V)} =  h_2^{(V)}  - \frac{2}{9}\,\frac{H^2}{(M+ \Delta)^2}\,, 
\nonumber \\
&& \bar h_3^{(V)} = h_3^{(V)}  + \frac{4}{27}\,\frac{H^2}{(M+ \Delta)^2} -  \frac{1}{6}\,\frac{\beta_{6}\,C^2}{M\,(M+\Delta)}\,,
\nonumber\\
&& \bar h_n^{(S)} = h_n^{(S)} \quad {\rm for} \qquad n \neq 5 \qquad {\rm and }\qquad \bar h_5^{(S)} = h_5^{(S)} -   \frac{C^2}{3\,M}\,\beta_5\,,
\label{Q4-renormalization}
\end{eqnarray}
where the coefficients $\alpha_{5,6}$ and $\beta_{5,6}$ depend on the ratio $\Delta/M$ only and approach $ 1$ in the limit with $\Delta \to 0$. They are detailed in Appendix A and Appendix B of \cite{Lutz:2018cqo}. 
The renormalization scale dependence of $c_n$ and $e_n$ is determined from (\ref{res-running}) with  
\allowdisplaybreaks[1]
\begin{eqnarray}
&& \Gamma_{c_0} =   \frac{20}{3}\,b_0+ 4\,b_D  -
\frac{1}{36}\, \Big( 30 \,\bar g_0^{(S)}+9 \,\bar g_1^{(S)}+26 \,\bar g_D^{(S)}\Big) 
\nonumber\\
&& \qquad \qquad -\, \frac{M}{144}\, \Big(30\,
   \bar g_0^{(V)}+9 \,\bar g_1^{(V)}+26 \,\bar g_D^{(V)}\Big) \,,
\nonumber\\
&& \Gamma_{c_1} = -
\frac{1}{24} \,\Big(4 \,\bar g_1^{(S)}+\bar g_1^{(V)}\, M \Big)\,,
\nonumber\\
&& \Gamma_{c_2} =  \frac{2}{3}\,b_D + \frac{1}{16} \,\Big(4\, (\bar g_1^{(S)}+\bar g_D^{(S)})+M\, (\bar g_1^{(V)}+\bar g_D^{(V)})\Big) \,,
\nonumber\\
&& \Gamma_{c_3} = \frac{2}{3}\,b_F  +\frac{1}{16} \,\Big(4\, \bar g_F^{(S)}+\bar g_F^{(V)}\, M\Big ) \,,
\nonumber\\
&& \Gamma_{c_4} = \frac{44}{9}\,b_D-\frac{1}{72} \,\Big( 36\, \bar g_1^{(S)}+52 \,\bar g_D^{(S)} + M\,( 9\, \bar g_1^{(V)} + 13\, \bar g_D^{(V)})
  \Big)\,,
\nonumber\\
&& \Gamma_{c_5} = \frac{44}{9}\,b_F-\frac{13}{72} \,\Big(4\, \bar g_F^{(S)}+\bar g_F^{(V)}\, M\Big) \,,
\nonumber\\
&& \Gamma_{c_6} = \frac{44}{9}\,b_0+ \frac{1}{432}\,\Big(-264 \,\bar g_0^{(S)}+108\, \bar g_1^{(S)} +32\, \bar g_D^{(S)} 
\nonumber\\
&& \qquad \qquad 
+ \, M\,\big(-66 \,\bar g_0^{(V)}+27\, \bar g_1^{(V)}  +8\, \bar g_D^{(V)} \big)\Big)\,,
\label{res-Gamma-ci} 
\\ \nonumber\\
&& \Gamma_{e_0} = \frac{20}{3}\,d_0+2\,d_D - \frac{1}{18}\,\Big(15\, \tilde h_1^{(S)}+13 \,\tilde h_2^{(S)}+9\, \tilde h_3^{(S) }
\Big)
\nonumber\\
&& \qquad -\, \frac{1}{72}\,( M+\Delta ) \,\Big( 15 \,\bar h_1^{(V)}+13\,\bar  h_2^{(V)} +9\,  \bar h_3^{(V)} \Big)  \,,
\nonumber\\
&& \Gamma_{e _1} = -\frac{1}{3} \,\tilde h_3^{(S)} - \frac{1}{12} \,( M+\Delta ) \,\bar h_3^{(V)}  \,,
\nonumber\\
&& \Gamma_{e_2} =  \frac{2}{3}\,d_D +  \frac{1}{2}\,\Big(\tilde h_2^{(S)}+\tilde h_3^{(S)}\Big)+
\frac{1}{8}\,( M+\Delta )\, \Big( \bar h_2^{(V)}+ \bar h_3^{(V)} \Big)\,,
\nonumber\\
&& \Gamma_{e_3} =   \frac{44 }{9}\,d_D -\frac{1}{9} \,\Big( 13 \,\tilde h_2^{(S)}+ 9\, \tilde h_3^{(S)}\Big) 
- \frac{1}{36} \,( M+\Delta )\,\Big( 13\, \bar h_2^{(V)} +9\,   \bar h_3^{(V)} \Big)\,,
\nonumber\\
&& \Gamma_{e_4} = \frac{44 }{9}\,d_0 +\frac{1}{54}\,\Big(-33 \,\tilde h_1^{(S)}+4 \,\tilde h_2^{(S)} \Big)
\nonumber\\
&& \qquad \qquad -\,\frac{1}{216}\,( M+\Delta )\,\Big( 33 \,\bar h_1^{(V)} -4\,\bar h_2^{(V)} \Big)\, ,
\nonumber\\
&&  \tilde h_1^{(S)} = \bar h_1^{(S)} + \frac{1}{3}\, \bar h_2^{(S)}\,, \qquad \qquad \tilde  h_1^{(V)} =  \bar h_1^{(V)} -  \frac{\bar h_2^{(S)}\,}{3\,(M + \Delta)} \,,
\nonumber\\
&&  \tilde h_2^{(S)} = \bar h_3^{(S)} + \frac{1}{3}\,\bar h_4^{(S)} \,, \qquad  \qquad\tilde  h_2^{(V)} =  \bar h_2^{(V)} -  \frac{\bar h_4^{(S)}\,}{3\,(M + \Delta)}\,,\quad 
\nonumber\\
&& \tilde h_3^{(S)} =\bar h_5^{(S)} +\frac{1}{3}\, \bar h_6^{(S)} \, ,\qquad \qquad \tilde  h_3^{(V)} =  \bar h_3^{(V)} -  \frac{\bar h_6^{(S)}\,}{3\,(M + \Delta)} \,.
\label{res-Gamma-ei}
\end{eqnarray}

\bibliography{literature}

\begin{thebibliography}{53}%
\makeatletter
\providecommand \@ifxundefined [1]{%
 \@ifx{#1\undefined}
}%
\providecommand \@ifnum [1]{%
 \ifnum #1\expandafter \@firstoftwo
 \else \expandafter \@secondoftwo
 \fi
}%
\providecommand \@ifx [1]{%
 \ifx #1\expandafter \@firstoftwo
 \else \expandafter \@secondoftwo
 \fi
}%
\providecommand \natexlab [1]{#1}%
\providecommand \enquote  [1]{``#1''}%
\providecommand \bibnamefont  [1]{#1}%
\providecommand \bibfnamefont [1]{#1}%
\providecommand \citenamefont [1]{#1}%
\providecommand \href@noop [0]{\@secondoftwo}%
\providecommand \href [0]{\begingroup \@sanitize@url \@href}%
\providecommand \@href[1]{\@@startlink{#1}\@@href}%
\providecommand \@@href[1]{\endgroup#1\@@endlink}%
\providecommand \@sanitize@url [0]{\catcode `\\12\catcode `\$12\catcode
  `\&12\catcode `\#12\catcode `\^12\catcode `\_12\catcode `\%12\relax}%
\providecommand \@@startlink[1]{}%
\providecommand \@@endlink[0]{}%
\providecommand \url  [0]{\begingroup\@sanitize@url \@url }%
\providecommand \@url [1]{\endgroup\@href {#1}{\urlprefix }}%
\providecommand \urlprefix  [0]{URL }%
\providecommand \Eprint [0]{\href }%
\providecommand \doibase [0]{http://dx.doi.org/}%
\providecommand \selectlanguage [0]{\@gobble}%
\providecommand \bibinfo  [0]{\@secondoftwo}%
\providecommand \bibfield  [0]{\@secondoftwo}%
\providecommand \translation [1]{[#1]}%
\providecommand \BibitemOpen [0]{}%
\providecommand \bibitemStop [0]{}%
\providecommand \bibitemNoStop [0]{.\EOS\space}%
\providecommand \EOS [0]{\spacefactor3000\relax}%
\providecommand \BibitemShut  [1]{\csname bibitem#1\endcsname}%
\let\auto@bib@innerbib\@empty
\bibitem [{\citenamefont {Guo}\ \emph {et~al.}(2018)\citenamefont {Guo},
  \citenamefont {Heo},\ and\ \citenamefont {Lutz}}]{Guo:2018kno}%
  \BibitemOpen
  \bibfield  {author} {\bibinfo {author} {\bibfnamefont {X.-Y.}\ \bibnamefont
  {Guo}}, \bibinfo {author} {\bibfnamefont {Y.}~\bibnamefont {Heo}}, \ and\
  \bibinfo {author} {\bibfnamefont {M.~F.~M.}\ \bibnamefont {Lutz}},\ }\href
  {\doibase 10.1103/PhysRevD.98.014510} {\bibfield  {journal} {\bibinfo
  {journal} {Phys. Rev. D}\ }\textbf {\bibinfo {volume} {98}},\ \bibinfo
  {pages} {014510} (\bibinfo {year} {2018})},\ \Eprint
  {http://arxiv.org/abs/1801.10122} {arXiv:1801.10122 [hep-lat]} \BibitemShut
  {NoStop}%
\bibitem [{\citenamefont {Lutz}\ \emph {et~al.}(2022)\citenamefont {Lutz},
  \citenamefont {Guo}, \citenamefont {Heo},\ and\ \citenamefont
  {Korpa}}]{Lutz:2022enz}%
  \BibitemOpen
  \bibfield  {author} {\bibinfo {author} {\bibfnamefont {M.~F.~M.}\
  \bibnamefont {Lutz}}, \bibinfo {author} {\bibfnamefont {X.-Y.}\ \bibnamefont
  {Guo}}, \bibinfo {author} {\bibfnamefont {Y.}~\bibnamefont {Heo}}, \ and\
  \bibinfo {author} {\bibfnamefont {C.~L.}\ \bibnamefont {Korpa}},\ }\href
  {\doibase 10.1103/PhysRevD.106.114038} {\bibfield  {journal} {\bibinfo
  {journal} {Phys. Rev. D}\ }\textbf {\bibinfo {volume} {106}},\ \bibinfo
  {pages} {114038} (\bibinfo {year} {2022})},\ \Eprint
  {http://arxiv.org/abs/2209.10601} {arXiv:2209.10601 [hep-ph]} \BibitemShut
  {NoStop}%
\bibitem [{\citenamefont {Korpa}\ \emph {et~al.}(2022)\citenamefont {Korpa},
  \citenamefont {Lutz}, \citenamefont {Guo},\ and\ \citenamefont
  {Heo}}]{Korpa:2022voo}%
  \BibitemOpen
  \bibfield  {author} {\bibinfo {author} {\bibfnamefont {C.~L.}\ \bibnamefont
  {Korpa}}, \bibinfo {author} {\bibfnamefont {M.~F.~M.}\ \bibnamefont {Lutz}},
  \bibinfo {author} {\bibfnamefont {X.-Y.}\ \bibnamefont {Guo}}, \ and\
  \bibinfo {author} {\bibfnamefont {Y.}~\bibnamefont {Heo}},\ }\href@noop {} {\
   (\bibinfo {year} {2022})},\ \Eprint {http://arxiv.org/abs/2211.03508}
  {arXiv:2211.03508 [hep-ph]} \BibitemShut {NoStop}%
\bibitem [{\citenamefont {Kalinowski}\ and\ \citenamefont
  {Wagner}(2015)}]{Kalinowski:2015bwa}%
  \BibitemOpen
  \bibfield  {author} {\bibinfo {author} {\bibfnamefont {M.}~\bibnamefont
  {Kalinowski}}\ and\ \bibinfo {author} {\bibfnamefont {M.}~\bibnamefont
  {Wagner}},\ }\href {\doibase 10.1103/PhysRevD.92.094508} {\bibfield
  {journal} {\bibinfo  {journal} {Phys. Rev.}\ }\textbf {\bibinfo {volume}
  {D92}},\ \bibinfo {pages} {094508} (\bibinfo {year} {2015})},\ \Eprint
  {http://arxiv.org/abs/1509.02396} {arXiv:1509.02396 [hep-lat]} \BibitemShut
  {NoStop}%
\bibitem [{\citenamefont {Na}\ \emph {et~al.}(2012)\citenamefont {Na},
  \citenamefont {Davies}, \citenamefont {Follana}, \citenamefont {Lepage},\
  and\ \citenamefont {Shigemitsu}}]{Na:2012iu}%
  \BibitemOpen
  \bibfield  {author} {\bibinfo {author} {\bibfnamefont {H.}~\bibnamefont
  {Na}}, \bibinfo {author} {\bibfnamefont {C.~T.}\ \bibnamefont {Davies}},
  \bibinfo {author} {\bibfnamefont {E.}~\bibnamefont {Follana}}, \bibinfo
  {author} {\bibfnamefont {G.~P.}\ \bibnamefont {Lepage}}, \ and\ \bibinfo
  {author} {\bibfnamefont {J.}~\bibnamefont {Shigemitsu}},\ }\href {\doibase
  10.1103/PhysRevD.86.054510} {\bibfield  {journal} {\bibinfo  {journal}
  {Phys.Rev.}\ }\textbf {\bibinfo {volume} {D86}},\ \bibinfo {pages} {054510}
  (\bibinfo {year} {2012})},\ \Eprint {http://arxiv.org/abs/1206.4936}
  {arXiv:1206.4936 [hep-lat]} \BibitemShut {NoStop}%
\bibitem [{\citenamefont {Moir}\ \emph {et~al.}(2016)\citenamefont {Moir},
  \citenamefont {Peardon}, \citenamefont {Ryan}, \citenamefont {Thomas},\ and\
  \citenamefont {Wilson}}]{Moir:2016srx}%
  \BibitemOpen
  \bibfield  {author} {\bibinfo {author} {\bibfnamefont {G.}~\bibnamefont
  {Moir}}, \bibinfo {author} {\bibfnamefont {M.}~\bibnamefont {Peardon}},
  \bibinfo {author} {\bibfnamefont {S.~M.}\ \bibnamefont {Ryan}}, \bibinfo
  {author} {\bibfnamefont {C.~E.}\ \bibnamefont {Thomas}}, \ and\ \bibinfo
  {author} {\bibfnamefont {D.~J.}\ \bibnamefont {Wilson}},\ }\href {\doibase
  10.1007/JHEP10(2016)011} {\bibfield  {journal} {\bibinfo  {journal} {JHEP}\
  }\textbf {\bibinfo {volume} {10}},\ \bibinfo {pages} {011} (\bibinfo {year}
  {2016})},\ \Eprint {http://arxiv.org/abs/1607.07093} {arXiv:1607.07093
  [hep-lat]} \BibitemShut {NoStop}%
\bibitem [{\citenamefont {Cheung}\ \emph {et~al.}(2021)\citenamefont {Cheung},
  \citenamefont {Thomas}, \citenamefont {Wilson}, \citenamefont {Moir},
  \citenamefont {Peardon},\ and\ \citenamefont {Ryan}}]{Cheung:2020mql}%
  \BibitemOpen
  \bibfield  {author} {\bibinfo {author} {\bibfnamefont {G.~K.~C.}\
  \bibnamefont {Cheung}}, \bibinfo {author} {\bibfnamefont {C.~E.}\
  \bibnamefont {Thomas}}, \bibinfo {author} {\bibfnamefont {D.~J.}\
  \bibnamefont {Wilson}}, \bibinfo {author} {\bibfnamefont {G.}~\bibnamefont
  {Moir}}, \bibinfo {author} {\bibfnamefont {M.}~\bibnamefont {Peardon}}, \
  and\ \bibinfo {author} {\bibfnamefont {S.~M.}\ \bibnamefont {Ryan}} (\bibinfo
  {collaboration} {Hadron Spectrum}),\ }\href {\doibase
  10.1007/JHEP02(2021)100} {\bibfield  {journal} {\bibinfo  {journal} {JHEP}\
  }\textbf {\bibinfo {volume} {02}},\ \bibinfo {pages} {100} (\bibinfo {year}
  {2021})},\ \Eprint {http://arxiv.org/abs/2008.06432} {arXiv:2008.06432
  [hep-lat]} \BibitemShut {NoStop}%
\bibitem [{\citenamefont {Gayer}\ \emph {et~al.}(2021)\citenamefont {Gayer},
  \citenamefont {Lang}, \citenamefont {Ryan}, \citenamefont {Tims},
  \citenamefont {Thomas},\ and\ \citenamefont {Wilson}}]{Gayer:2021xzv}%
  \BibitemOpen
  \bibfield  {author} {\bibinfo {author} {\bibfnamefont {L.}~\bibnamefont
  {Gayer}}, \bibinfo {author} {\bibfnamefont {N.}~\bibnamefont {Lang}},
  \bibinfo {author} {\bibfnamefont {S.~M.}\ \bibnamefont {Ryan}}, \bibinfo
  {author} {\bibfnamefont {D.}~\bibnamefont {Tims}}, \bibinfo {author}
  {\bibfnamefont {C.~E.}\ \bibnamefont {Thomas}}, \ and\ \bibinfo {author}
  {\bibfnamefont {D.~J.}\ \bibnamefont {Wilson}} (\bibinfo {collaboration}
  {Hadron Spectrum}),\ }\href {\doibase 10.1007/JHEP07(2021)123} {\bibfield
  {journal} {\bibinfo  {journal} {JHEP}\ }\textbf {\bibinfo {volume} {07}},\
  \bibinfo {pages} {123} (\bibinfo {year} {2021})},\ \Eprint
  {http://arxiv.org/abs/2102.04973} {arXiv:2102.04973 [hep-lat]} \BibitemShut
  {NoStop}%
\bibitem [{\citenamefont {Bali}\ \emph {et~al.}(2022)\citenamefont {Bali},
  \citenamefont {Collins}, \citenamefont {Georg}, \citenamefont {Jenkins},
  \citenamefont {Korcyl}, \citenamefont {Sch\"afer}, \citenamefont {Scholz},
  \citenamefont {Simeth}, \citenamefont {S\"oldner},\ and\ \citenamefont
  {Weish\"aupl}}]{RQCD:2022xux}%
  \BibitemOpen
  \bibfield  {author} {\bibinfo {author} {\bibfnamefont {G.~S.}\ \bibnamefont
  {Bali}}, \bibinfo {author} {\bibfnamefont {S.}~\bibnamefont {Collins}},
  \bibinfo {author} {\bibfnamefont {P.}~\bibnamefont {Georg}}, \bibinfo
  {author} {\bibfnamefont {D.}~\bibnamefont {Jenkins}}, \bibinfo {author}
  {\bibfnamefont {P.}~\bibnamefont {Korcyl}}, \bibinfo {author} {\bibfnamefont
  {A.}~\bibnamefont {Sch\"afer}}, \bibinfo {author} {\bibfnamefont {E.~E.}\
  \bibnamefont {Scholz}}, \bibinfo {author} {\bibfnamefont {J.}~\bibnamefont
  {Simeth}}, \bibinfo {author} {\bibfnamefont {W.}~\bibnamefont {S\"oldner}}, \
  and\ \bibinfo {author} {\bibfnamefont {S.}~\bibnamefont {Weish\"aupl}}
  (\bibinfo {collaboration} {RQCD}),\ }\href@noop {} {\  (\bibinfo {year}
  {2022})},\ \Eprint {http://arxiv.org/abs/2211.03744} {arXiv:2211.03744
  [hep-lat]} \BibitemShut {NoStop}%
\bibitem [{\citenamefont {Lutz}\ and\ \citenamefont
  {Semke}(2012)}]{Lutz:2012mq}%
  \BibitemOpen
  \bibfield  {author} {\bibinfo {author} {\bibfnamefont {M.~F.~M.}\
  \bibnamefont {Lutz}}\ and\ \bibinfo {author} {\bibfnamefont {A.}~\bibnamefont
  {Semke}},\ }\href {\doibase 10.1103/PhysRevD.86.091502} {\bibfield  {journal}
  {\bibinfo  {journal} {Phys. Rev. D}\ }\textbf {\bibinfo {volume} {86}},\
  \bibinfo {pages} {091502} (\bibinfo {year} {2012})},\ \Eprint
  {http://arxiv.org/abs/1209.2791} {arXiv:1209.2791 [hep-lat]} \BibitemShut
  {NoStop}%
\bibitem [{\citenamefont {Lutz}\ \emph {et~al.}(2014)\citenamefont {Lutz},
  \citenamefont {Bavontaweepanya}, \citenamefont {Kobdaj},\ and\ \citenamefont
  {Schwarz}}]{Lutz:2014oxa}%
  \BibitemOpen
  \bibfield  {author} {\bibinfo {author} {\bibfnamefont {M.~F.~M.}\
  \bibnamefont {Lutz}}, \bibinfo {author} {\bibfnamefont {R.}~\bibnamefont
  {Bavontaweepanya}}, \bibinfo {author} {\bibfnamefont {C.}~\bibnamefont
  {Kobdaj}}, \ and\ \bibinfo {author} {\bibfnamefont {K.}~\bibnamefont
  {Schwarz}},\ }\href {\doibase 10.1103/PhysRevD.90.054505} {\bibfield
  {journal} {\bibinfo  {journal} {Phys. Rev.}\ }\textbf {\bibinfo {volume}
  {D90}},\ \bibinfo {pages} {054505} (\bibinfo {year} {2014})},\ \Eprint
  {http://arxiv.org/abs/1401.7805} {arXiv:1401.7805 [hep-lat]} \BibitemShut
  {NoStop}%
\bibitem [{\citenamefont {Lutz}\ \emph {et~al.}(2018)\citenamefont {Lutz},
  \citenamefont {Heo},\ and\ \citenamefont {Guo}}]{Lutz:2018cqo}%
  \BibitemOpen
  \bibfield  {author} {\bibinfo {author} {\bibfnamefont {M.~F.~M.}\
  \bibnamefont {Lutz}}, \bibinfo {author} {\bibfnamefont {Y.}~\bibnamefont
  {Heo}}, \ and\ \bibinfo {author} {\bibfnamefont {X.-Y.}\ \bibnamefont
  {Guo}},\ }\href {\doibase 10.1016/j.nuclphysa.2018.05.007} {\bibfield
  {journal} {\bibinfo  {journal} {Nucl. Phys. A}\ }\textbf {\bibinfo {volume}
  {977}},\ \bibinfo {pages} {146} (\bibinfo {year} {2018})},\ \Eprint
  {http://arxiv.org/abs/1801.06417} {arXiv:1801.06417 [hep-lat]} \BibitemShut
  {NoStop}%
\bibitem [{\citenamefont {Guo}\ \emph {et~al.}(2020)\citenamefont {Guo},
  \citenamefont {Heo},\ and\ \citenamefont {Lutz}}]{Guo:2019nyp}%
  \BibitemOpen
  \bibfield  {author} {\bibinfo {author} {\bibfnamefont {X.-Y.}\ \bibnamefont
  {Guo}}, \bibinfo {author} {\bibfnamefont {Y.}~\bibnamefont {Heo}}, \ and\
  \bibinfo {author} {\bibfnamefont {M.~F.~M.}\ \bibnamefont {Lutz}},\ }\href
  {\doibase 10.1140/epjc/s10052-020-7818-9} {\bibfield  {journal} {\bibinfo
  {journal} {Eur. Phys. J. C}\ }\textbf {\bibinfo {volume} {80}},\ \bibinfo
  {pages} {260} (\bibinfo {year} {2020})},\ \Eprint
  {http://arxiv.org/abs/1907.00714} {arXiv:1907.00714 [hep-lat]} \BibitemShut
  {NoStop}%
\bibitem [{\citenamefont {Aoki}\ \emph {et~al.}(2009)\citenamefont {Aoki} \emph
  {et~al.}}]{PACS-CS:2008bkb}%
  \BibitemOpen
  \bibfield  {author} {\bibinfo {author} {\bibfnamefont {S.}~\bibnamefont
  {Aoki}} \emph {et~al.} (\bibinfo {collaboration} {PACS-CS}),\ }\href
  {\doibase 10.1103/PhysRevD.79.034503} {\bibfield  {journal} {\bibinfo
  {journal} {Phys. Rev. D}\ }\textbf {\bibinfo {volume} {79}},\ \bibinfo
  {pages} {034503} (\bibinfo {year} {2009})},\ \Eprint
  {http://arxiv.org/abs/0807.1661} {arXiv:0807.1661 [hep-lat]} \BibitemShut
  {NoStop}%
\bibitem [{\citenamefont {Lin}\ \emph {et~al.}(2009)\citenamefont {Lin} \emph
  {et~al.}}]{HadronSpectrum:2008xlg}%
  \BibitemOpen
  \bibfield  {author} {\bibinfo {author} {\bibfnamefont {H.-W.}\ \bibnamefont
  {Lin}} \emph {et~al.} (\bibinfo {collaboration} {Hadron Spectrum}),\ }\href
  {\doibase 10.1103/PhysRevD.79.034502} {\bibfield  {journal} {\bibinfo
  {journal} {Phys. Rev. D}\ }\textbf {\bibinfo {volume} {79}},\ \bibinfo
  {pages} {034502} (\bibinfo {year} {2009})},\ \Eprint
  {http://arxiv.org/abs/0810.3588} {arXiv:0810.3588 [hep-lat]} \BibitemShut
  {NoStop}%
\bibitem [{\citenamefont {Walker-Loud}\ \emph {et~al.}(2009)\citenamefont
  {Walker-Loud} \emph {et~al.}}]{Walker-Loud:2008rui}%
  \BibitemOpen
  \bibfield  {author} {\bibinfo {author} {\bibfnamefont {A.}~\bibnamefont
  {Walker-Loud}} \emph {et~al.},\ }\href {\doibase 10.1103/PhysRevD.79.054502}
  {\bibfield  {journal} {\bibinfo  {journal} {Phys. Rev. D}\ }\textbf {\bibinfo
  {volume} {79}},\ \bibinfo {pages} {054502} (\bibinfo {year} {2009})},\
  \Eprint {http://arxiv.org/abs/0806.4549} {arXiv:0806.4549 [hep-lat]}
  \BibitemShut {NoStop}%
\bibitem [{\citenamefont {Beane}\ \emph {et~al.}(2011)\citenamefont {Beane},
  \citenamefont {Chang}, \citenamefont {Detmold}, \citenamefont {Lin},
  \citenamefont {Luu}, \citenamefont {Orginos}, \citenamefont {Parreno},
  \citenamefont {Savage}, \citenamefont {Torok},\ and\ \citenamefont
  {Walker-Loud}}]{Beane:2011pc}%
  \BibitemOpen
  \bibfield  {author} {\bibinfo {author} {\bibfnamefont {S.~R.}\ \bibnamefont
  {Beane}}, \bibinfo {author} {\bibfnamefont {E.}~\bibnamefont {Chang}},
  \bibinfo {author} {\bibfnamefont {W.}~\bibnamefont {Detmold}}, \bibinfo
  {author} {\bibfnamefont {H.~W.}\ \bibnamefont {Lin}}, \bibinfo {author}
  {\bibfnamefont {T.~C.}\ \bibnamefont {Luu}}, \bibinfo {author} {\bibfnamefont
  {K.}~\bibnamefont {Orginos}}, \bibinfo {author} {\bibfnamefont
  {A.}~\bibnamefont {Parreno}}, \bibinfo {author} {\bibfnamefont {M.~J.}\
  \bibnamefont {Savage}}, \bibinfo {author} {\bibfnamefont {A.}~\bibnamefont
  {Torok}}, \ and\ \bibinfo {author} {\bibfnamefont {A.}~\bibnamefont
  {Walker-Loud}},\ }\href {\doibase 10.1103/PhysRevD.84.014507} {\bibfield
  {journal} {\bibinfo  {journal} {Phys. Rev. D}\ }\textbf {\bibinfo {volume}
  {84}},\ \bibinfo {pages} {014507} (\bibinfo {year} {2011})},\ \Eprint
  {http://arxiv.org/abs/1104.4101} {arXiv:1104.4101 [hep-lat]} \BibitemShut
  {NoStop}%
\bibitem [{\citenamefont {Bietenholz}\ \emph {et~al.}(2011)\citenamefont
  {Bietenholz} \emph {et~al.}}]{Bietenholz:2011qq}%
  \BibitemOpen
  \bibfield  {author} {\bibinfo {author} {\bibfnamefont {W.}~\bibnamefont
  {Bietenholz}} \emph {et~al.},\ }\href {\doibase 10.1103/PhysRevD.84.054509}
  {\bibfield  {journal} {\bibinfo  {journal} {Phys. Rev. D}\ }\textbf {\bibinfo
  {volume} {84}},\ \bibinfo {pages} {054509} (\bibinfo {year} {2011})},\
  \Eprint {http://arxiv.org/abs/1102.5300} {arXiv:1102.5300 [hep-lat]}
  \BibitemShut {NoStop}%
\bibitem [{\citenamefont {Alexandrou}\ \emph {et~al.}(2013)\citenamefont
  {Alexandrou}, \citenamefont {Constantinou}, \citenamefont {Dinter},
  \citenamefont {Drach}, \citenamefont {Jansen}, \citenamefont {Kallidonis},\
  and\ \citenamefont {Koutsou}}]{Alexandrou:2013joa}%
  \BibitemOpen
  \bibfield  {author} {\bibinfo {author} {\bibfnamefont {C.}~\bibnamefont
  {Alexandrou}}, \bibinfo {author} {\bibfnamefont {M.}~\bibnamefont
  {Constantinou}}, \bibinfo {author} {\bibfnamefont {S.}~\bibnamefont
  {Dinter}}, \bibinfo {author} {\bibfnamefont {V.}~\bibnamefont {Drach}},
  \bibinfo {author} {\bibfnamefont {K.}~\bibnamefont {Jansen}}, \bibinfo
  {author} {\bibfnamefont {C.}~\bibnamefont {Kallidonis}}, \ and\ \bibinfo
  {author} {\bibfnamefont {G.}~\bibnamefont {Koutsou}},\ }\href {\doibase
  10.1103/PhysRevD.88.014509} {\bibfield  {journal} {\bibinfo  {journal} {Phys.
  Rev. D}\ }\textbf {\bibinfo {volume} {88}},\ \bibinfo {pages} {014509}
  (\bibinfo {year} {2013})},\ \Eprint {http://arxiv.org/abs/1303.5979}
  {arXiv:1303.5979 [hep-lat]} \BibitemShut {NoStop}%
\bibitem [{\citenamefont {Lutz}\ and\ \citenamefont
  {Semke}(2011)}]{Lutz:2010se}%
  \BibitemOpen
  \bibfield  {author} {\bibinfo {author} {\bibfnamefont {M.~F.~M.}\
  \bibnamefont {Lutz}}\ and\ \bibinfo {author} {\bibfnamefont {A.}~\bibnamefont
  {Semke}},\ }\href {\doibase 10.1103/PhysRevD.83.034008} {\bibfield  {journal}
  {\bibinfo  {journal} {Phys. Rev. D}\ }\textbf {\bibinfo {volume} {83}},\
  \bibinfo {pages} {034008} (\bibinfo {year} {2011})},\ \Eprint
  {http://arxiv.org/abs/1012.4365} {arXiv:1012.4365 [hep-ph]} \BibitemShut
  {NoStop}%
\bibitem [{\citenamefont {Gasser}\ \emph {et~al.}(1988)\citenamefont {Gasser},
  \citenamefont {Sainio},\ and\ \citenamefont {Svarc}}]{Gasser:1987rb}%
  \BibitemOpen
  \bibfield  {author} {\bibinfo {author} {\bibfnamefont {J.}~\bibnamefont
  {Gasser}}, \bibinfo {author} {\bibfnamefont {M.~E.}\ \bibnamefont {Sainio}},
  \ and\ \bibinfo {author} {\bibfnamefont {A.}~\bibnamefont {Svarc}},\ }\href
  {\doibase 10.1016/0550-3213(88)90108-3} {\bibfield  {journal} {\bibinfo
  {journal} {Nucl. Phys. B}\ }\textbf {\bibinfo {volume} {307}},\ \bibinfo
  {pages} {779} (\bibinfo {year} {1988})}\BibitemShut {NoStop}%
\bibitem [{\citenamefont {Jenkins}\ and\ \citenamefont
  {Manohar}(1991)}]{Jenkins:1990jv}%
  \BibitemOpen
  \bibfield  {author} {\bibinfo {author} {\bibfnamefont {E.~E.}\ \bibnamefont
  {Jenkins}}\ and\ \bibinfo {author} {\bibfnamefont {A.~V.}\ \bibnamefont
  {Manohar}},\ }\href {\doibase 10.1016/0370-2693(91)90266-S} {\bibfield
  {journal} {\bibinfo  {journal} {Phys. Lett. B}\ }\textbf {\bibinfo {volume}
  {255}},\ \bibinfo {pages} {558} (\bibinfo {year} {1991})}\BibitemShut
  {NoStop}%
\bibitem [{\citenamefont {Krause}(1990)}]{Krause:1990xc}%
  \BibitemOpen
  \bibfield  {author} {\bibinfo {author} {\bibfnamefont {A.}~\bibnamefont
  {Krause}},\ }\href {\doibase 10.5169/seals-116214} {\bibfield  {journal}
  {\bibinfo  {journal} {Helv. Phys. Acta}\ }\textbf {\bibinfo {volume} {63}},\
  \bibinfo {pages} {3} (\bibinfo {year} {1990})}\BibitemShut {NoStop}%
\bibitem [{\citenamefont {Becher}\ and\ \citenamefont
  {Leutwyler}(1999)}]{Becher:1999he}%
  \BibitemOpen
  \bibfield  {author} {\bibinfo {author} {\bibfnamefont {T.}~\bibnamefont
  {Becher}}\ and\ \bibinfo {author} {\bibfnamefont {H.}~\bibnamefont
  {Leutwyler}},\ }\href {\doibase 10.1007/PL00021673} {\bibfield  {journal}
  {\bibinfo  {journal} {Eur. Phys. J. C}\ }\textbf {\bibinfo {volume} {9}},\
  \bibinfo {pages} {643} (\bibinfo {year} {1999})},\ \Eprint
  {http://arxiv.org/abs/hep-ph/9901384} {arXiv:hep-ph/9901384} \BibitemShut
  {NoStop}%
\bibitem [{\citenamefont {Lutz}\ and\ \citenamefont
  {Kolomeitsev}(2002)}]{Lutz:2001yb}%
  \BibitemOpen
  \bibfield  {author} {\bibinfo {author} {\bibfnamefont {M.~F.~M.}\
  \bibnamefont {Lutz}}\ and\ \bibinfo {author} {\bibfnamefont {E.~E.}\
  \bibnamefont {Kolomeitsev}},\ }\href {\doibase 10.1016/S0375-9474(01)01312-4}
  {\bibfield  {journal} {\bibinfo  {journal} {Nucl. Phys. A}\ }\textbf
  {\bibinfo {volume} {700}},\ \bibinfo {pages} {193} (\bibinfo {year}
  {2002})},\ \Eprint {http://arxiv.org/abs/nucl-th/0105042}
  {arXiv:nucl-th/0105042} \BibitemShut {NoStop}%
\bibitem [{\citenamefont {Scherer}(2003)}]{Scherer:2002tk}%
  \BibitemOpen
  \bibfield  {author} {\bibinfo {author} {\bibfnamefont {S.}~\bibnamefont
  {Scherer}},\ }\href@noop {} {\bibfield  {journal} {\bibinfo  {journal} {Adv.
  Nucl. Phys.}\ }\textbf {\bibinfo {volume} {27}},\ \bibinfo {pages} {277}
  (\bibinfo {year} {2003})},\ \Eprint {http://arxiv.org/abs/hep-ph/0210398}
  {arXiv:hep-ph/0210398} \BibitemShut {NoStop}%
\bibitem [{\citenamefont {Lehnhart}\ \emph {et~al.}(2005)\citenamefont
  {Lehnhart}, \citenamefont {Gegelia},\ and\ \citenamefont
  {Scherer}}]{Lehnhart:2004vi}%
  \BibitemOpen
  \bibfield  {author} {\bibinfo {author} {\bibfnamefont {B.~C.}\ \bibnamefont
  {Lehnhart}}, \bibinfo {author} {\bibfnamefont {J.}~\bibnamefont {Gegelia}}, \
  and\ \bibinfo {author} {\bibfnamefont {S.}~\bibnamefont {Scherer}},\ }\href
  {\doibase 10.1088/0954-3899/31/2/002} {\bibfield  {journal} {\bibinfo
  {journal} {J. Phys. G}\ }\textbf {\bibinfo {volume} {31}},\ \bibinfo {pages}
  {89} (\bibinfo {year} {2005})},\ \Eprint
  {http://arxiv.org/abs/hep-ph/0412092} {arXiv:hep-ph/0412092} \BibitemShut
  {NoStop}%
\bibitem [{\citenamefont {Semke}\ and\ \citenamefont
  {Lutz}(2006)}]{Semke:2005sn}%
  \BibitemOpen
  \bibfield  {author} {\bibinfo {author} {\bibfnamefont {A.}~\bibnamefont
  {Semke}}\ and\ \bibinfo {author} {\bibfnamefont {M.~F.~M.}\ \bibnamefont
  {Lutz}},\ }\href {\doibase 10.1016/j.nuclphysa.2006.07.043} {\bibfield
  {journal} {\bibinfo  {journal} {Nucl.Phys.}\ }\textbf {\bibinfo {volume}
  {A778}},\ \bibinfo {pages} {153} (\bibinfo {year} {2006})},\ \Eprint
  {http://arxiv.org/abs/nucl-th/0511061} {arXiv:nucl-th/0511061 [nucl-th]}
  \BibitemShut {NoStop}%
\bibitem [{\citenamefont {Hacker}\ \emph {et~al.}(2005)\citenamefont {Hacker},
  \citenamefont {Wies}, \citenamefont {Gegelia},\ and\ \citenamefont
  {Scherer}}]{Hacker:2005fh}%
  \BibitemOpen
  \bibfield  {author} {\bibinfo {author} {\bibfnamefont {C.}~\bibnamefont
  {Hacker}}, \bibinfo {author} {\bibfnamefont {N.}~\bibnamefont {Wies}},
  \bibinfo {author} {\bibfnamefont {J.}~\bibnamefont {Gegelia}}, \ and\
  \bibinfo {author} {\bibfnamefont {S.}~\bibnamefont {Scherer}},\ }\href
  {\doibase 10.1103/PhysRevC.72.055203} {\bibfield  {journal} {\bibinfo
  {journal} {Phys. Rev. C}\ }\textbf {\bibinfo {volume} {72}},\ \bibinfo
  {pages} {055203} (\bibinfo {year} {2005})},\ \Eprint
  {http://arxiv.org/abs/hep-ph/0505043} {arXiv:hep-ph/0505043} \BibitemShut
  {NoStop}%
\bibitem [{\citenamefont {Frink}\ and\ \citenamefont
  {Meissner}(2006)}]{Frink:2006hx}%
  \BibitemOpen
  \bibfield  {author} {\bibinfo {author} {\bibfnamefont {M.}~\bibnamefont
  {Frink}}\ and\ \bibinfo {author} {\bibfnamefont {U.-G.}\ \bibnamefont
  {Meissner}},\ }\href {\doibase 10.1140/epja/i2006-10105-x} {\bibfield
  {journal} {\bibinfo  {journal} {Eur. Phys. J. A}\ }\textbf {\bibinfo {volume}
  {29}},\ \bibinfo {pages} {255} (\bibinfo {year} {2006})},\ \Eprint
  {http://arxiv.org/abs/hep-ph/0609256} {arXiv:hep-ph/0609256} \BibitemShut
  {NoStop}%
\bibitem [{\citenamefont {Martin~Camalich}\ \emph {et~al.}(2010)\citenamefont
  {Martin~Camalich}, \citenamefont {Geng},\ and\ \citenamefont
  {Vicente~Vacas}}]{MartinCamalich:2010fp}%
  \BibitemOpen
  \bibfield  {author} {\bibinfo {author} {\bibfnamefont {J.}~\bibnamefont
  {Martin~Camalich}}, \bibinfo {author} {\bibfnamefont {L.~S.}\ \bibnamefont
  {Geng}}, \ and\ \bibinfo {author} {\bibfnamefont {M.~J.}\ \bibnamefont
  {Vicente~Vacas}},\ }\href {\doibase 10.1103/PhysRevD.82.074504} {\bibfield
  {journal} {\bibinfo  {journal} {Phys. Rev. D}\ }\textbf {\bibinfo {volume}
  {82}},\ \bibinfo {pages} {074504} (\bibinfo {year} {2010})},\ \Eprint
  {http://arxiv.org/abs/1003.1929} {arXiv:1003.1929 [hep-lat]} \BibitemShut
  {NoStop}%
\bibitem [{\citenamefont {Semke}\ and\ \citenamefont
  {Lutz}(2012{\natexlab{a}})}]{Semke:2011ez}%
  \BibitemOpen
  \bibfield  {author} {\bibinfo {author} {\bibfnamefont {A.}~\bibnamefont
  {Semke}}\ and\ \bibinfo {author} {\bibfnamefont {M.~F.~M.}\ \bibnamefont
  {Lutz}},\ }\href {\doibase 10.1103/PhysRevD.85.034001} {\bibfield  {journal}
  {\bibinfo  {journal} {Phys.Rev.}\ }\textbf {\bibinfo {volume} {D85}},\
  \bibinfo {pages} {034001} (\bibinfo {year} {2012}{\natexlab{a}})},\ \Eprint
  {http://arxiv.org/abs/1111.0238} {arXiv:1111.0238 [hep-ph]} \BibitemShut
  {NoStop}%
\bibitem [{\citenamefont {Semke}\ and\ \citenamefont
  {Lutz}(2012{\natexlab{b}})}]{Semke:2012gs}%
  \BibitemOpen
  \bibfield  {author} {\bibinfo {author} {\bibfnamefont {A.}~\bibnamefont
  {Semke}}\ and\ \bibinfo {author} {\bibfnamefont {M.~F.~M.}\ \bibnamefont
  {Lutz}},\ }\href {\doibase 10.1016/j.physletb.2012.09.008} {\bibfield
  {journal} {\bibinfo  {journal} {Phys. Lett. B}\ }\textbf {\bibinfo {volume}
  {717}},\ \bibinfo {pages} {242} (\bibinfo {year} {2012}{\natexlab{b}})},\
  \Eprint {http://arxiv.org/abs/1202.3556} {arXiv:1202.3556 [hep-ph]}
  \BibitemShut {NoStop}%
\bibitem [{\citenamefont {Ren}\ \emph {et~al.}(2012)\citenamefont {Ren},
  \citenamefont {Geng}, \citenamefont {Martin~Camalich}, \citenamefont {Meng},\
  and\ \citenamefont {Toki}}]{Ren:2012aj}%
  \BibitemOpen
  \bibfield  {author} {\bibinfo {author} {\bibfnamefont {X.~L.}\ \bibnamefont
  {Ren}}, \bibinfo {author} {\bibfnamefont {L.~S.}\ \bibnamefont {Geng}},
  \bibinfo {author} {\bibfnamefont {J.}~\bibnamefont {Martin~Camalich}},
  \bibinfo {author} {\bibfnamefont {J.}~\bibnamefont {Meng}}, \ and\ \bibinfo
  {author} {\bibfnamefont {H.}~\bibnamefont {Toki}},\ }\href {\doibase
  10.1007/JHEP12(2012)073} {\bibfield  {journal} {\bibinfo  {journal} {JHEP}\
  }\textbf {\bibinfo {volume} {12}},\ \bibinfo {pages} {073} (\bibinfo {year}
  {2012})},\ \Eprint {http://arxiv.org/abs/1209.3641} {arXiv:1209.3641
  [nucl-th]} \BibitemShut {NoStop}%
\bibitem [{\citenamefont {Ren}\ \emph {et~al.}(2014)\citenamefont {Ren},
  \citenamefont {Geng},\ and\ \citenamefont {Meng}}]{Ren:2013oaa}%
  \BibitemOpen
  \bibfield  {author} {\bibinfo {author} {\bibfnamefont {X.-L.}\ \bibnamefont
  {Ren}}, \bibinfo {author} {\bibfnamefont {L.-S.}\ \bibnamefont {Geng}}, \
  and\ \bibinfo {author} {\bibfnamefont {J.}~\bibnamefont {Meng}},\ }\href
  {\doibase 10.1103/PhysRevD.89.054034} {\bibfield  {journal} {\bibinfo
  {journal} {Phys. Rev. D}\ }\textbf {\bibinfo {volume} {89}},\ \bibinfo
  {pages} {054034} (\bibinfo {year} {2014})},\ \Eprint
  {http://arxiv.org/abs/1307.1896} {arXiv:1307.1896 [nucl-th]} \BibitemShut
  {NoStop}%
\bibitem [{\citenamefont {Holmberg}\ and\ \citenamefont
  {Leupold}(2018)}]{Holmberg:2018dtv}%
  \BibitemOpen
  \bibfield  {author} {\bibinfo {author} {\bibfnamefont {M.}~\bibnamefont
  {Holmberg}}\ and\ \bibinfo {author} {\bibfnamefont {S.}~\bibnamefont
  {Leupold}},\ }\href {\doibase 10.1140/epja/i2018-12533-3} {\bibfield
  {journal} {\bibinfo  {journal} {Eur. Phys. J. A}\ }\textbf {\bibinfo {volume}
  {54}},\ \bibinfo {pages} {103} (\bibinfo {year} {2018})},\ \Eprint
  {http://arxiv.org/abs/1802.05168} {arXiv:1802.05168 [hep-ph]} \BibitemShut
  {NoStop}%
\bibitem [{\citenamefont {Gasser}\ and\ \citenamefont
  {Leutwyler}(1985)}]{Gasser:1984gg}%
  \BibitemOpen
  \bibfield  {author} {\bibinfo {author} {\bibfnamefont {J.}~\bibnamefont
  {Gasser}}\ and\ \bibinfo {author} {\bibfnamefont {H.}~\bibnamefont
  {Leutwyler}},\ }\href {\doibase 10.1016/0550-3213(85)90492-4} {\bibfield
  {journal} {\bibinfo  {journal} {Nucl. Phys.}\ }\textbf {\bibinfo {volume}
  {B250}},\ \bibinfo {pages} {465} (\bibinfo {year} {1985})}\BibitemShut
  {NoStop}%
\bibitem [{\citenamefont {Bavontaweepanya}\ \emph {et~al.}(2018)\citenamefont
  {Bavontaweepanya}, \citenamefont {Guo},\ and\ \citenamefont
  {Lutz}}]{Bavontaweepanya:2018yds}%
  \BibitemOpen
  \bibfield  {author} {\bibinfo {author} {\bibfnamefont {R.}~\bibnamefont
  {Bavontaweepanya}}, \bibinfo {author} {\bibfnamefont {X.-Y.}\ \bibnamefont
  {Guo}}, \ and\ \bibinfo {author} {\bibfnamefont {M.~F.~M.}\ \bibnamefont
  {Lutz}},\ }\href {\doibase 10.1103/PhysRevD.98.056005} {\bibfield  {journal}
  {\bibinfo  {journal} {Phys. Rev. D}\ }\textbf {\bibinfo {volume} {98}},\
  \bibinfo {pages} {056005} (\bibinfo {year} {2018})},\ \Eprint
  {http://arxiv.org/abs/1801.10522} {arXiv:1801.10522 [hep-ph]} \BibitemShut
  {NoStop}%
\bibitem [{\citenamefont {Guo}\ and\ \citenamefont {Lutz}(2019)}]{Guo:2018zvl}%
  \BibitemOpen
  \bibfield  {author} {\bibinfo {author} {\bibfnamefont {X.-Y.}\ \bibnamefont
  {Guo}}\ and\ \bibinfo {author} {\bibfnamefont {M.~F.~M.}\ \bibnamefont
  {Lutz}},\ }\href {\doibase 10.1016/j.nuclphysa.2019.02.007} {\bibfield
  {journal} {\bibinfo  {journal} {Nucl. Phys. A}\ }\textbf {\bibinfo {volume}
  {988}},\ \bibinfo {pages} {48} (\bibinfo {year} {2019})},\ \Eprint
  {http://arxiv.org/abs/1810.07078} {arXiv:1810.07078 [hep-lat]} \BibitemShut
  {NoStop}%
\bibitem [{\citenamefont {Aoki}\ \emph {et~al.}(2021)\citenamefont {Aoki} \emph
  {et~al.}}]{Aoki:2021kgd}%
  \BibitemOpen
  \bibfield  {author} {\bibinfo {author} {\bibfnamefont {Y.}~\bibnamefont
  {Aoki}} \emph {et~al.},\ }\href@noop {} {\  (\bibinfo {year} {2021})},\
  \Eprint {http://arxiv.org/abs/2111.09849} {arXiv:2111.09849 [hep-lat]}
  \BibitemShut {NoStop}%
\bibitem [{\citenamefont {'t~Hooft}(1974)}]{tHooft:1973alw}%
  \BibitemOpen
  \bibfield  {author} {\bibinfo {author} {\bibfnamefont {G.}~\bibnamefont
  {'t~Hooft}},\ }\href {\doibase 10.1016/0550-3213(74)90154-0} {\bibfield
  {journal} {\bibinfo  {journal} {Nucl. Phys. B}\ }\textbf {\bibinfo {volume}
  {72}},\ \bibinfo {pages} {461} (\bibinfo {year} {1974})}\BibitemShut
  {NoStop}%
\bibitem [{\citenamefont {Witten}(1979)}]{Witten:1979kh}%
  \BibitemOpen
  \bibfield  {author} {\bibinfo {author} {\bibfnamefont {E.}~\bibnamefont
  {Witten}},\ }\href {\doibase 10.1016/0550-3213(79)90232-3} {\bibfield
  {journal} {\bibinfo  {journal} {Nucl. Phys. B}\ }\textbf {\bibinfo {volume}
  {160}},\ \bibinfo {pages} {57} (\bibinfo {year} {1979})}\BibitemShut
  {NoStop}%
\bibitem [{\citenamefont {Dashen}\ \emph {et~al.}(1994)\citenamefont {Dashen},
  \citenamefont {Jenkins},\ and\ \citenamefont {Manohar}}]{Dashen:1993jt}%
  \BibitemOpen
  \bibfield  {author} {\bibinfo {author} {\bibfnamefont {R.~F.}\ \bibnamefont
  {Dashen}}, \bibinfo {author} {\bibfnamefont {E.~E.}\ \bibnamefont {Jenkins}},
  \ and\ \bibinfo {author} {\bibfnamefont {A.~V.}\ \bibnamefont {Manohar}},\
  }\href {\doibase 10.1103/PhysRevD.51.2489} {\bibfield  {journal} {\bibinfo
  {journal} {Phys. Rev. D}\ }\textbf {\bibinfo {volume} {49}},\ \bibinfo
  {pages} {4713} (\bibinfo {year} {1994})},\ \bibinfo {note} {[Erratum:
  Phys.Rev.D 51, 2489 (1995)]},\ \Eprint {http://arxiv.org/abs/hep-ph/9310379}
  {arXiv:hep-ph/9310379} \BibitemShut {NoStop}%
\bibitem [{\citenamefont {Luty}\ and\ \citenamefont
  {March-Russell}(1994)}]{Luty:1993fu}%
  \BibitemOpen
  \bibfield  {author} {\bibinfo {author} {\bibfnamefont {M.~A.}\ \bibnamefont
  {Luty}}\ and\ \bibinfo {author} {\bibfnamefont {J.}~\bibnamefont
  {March-Russell}},\ }\href {\doibase 10.1016/0550-3213(94)90126-0} {\bibfield
  {journal} {\bibinfo  {journal} {Nucl. Phys. B}\ }\textbf {\bibinfo {volume}
  {426}},\ \bibinfo {pages} {71} (\bibinfo {year} {1994})},\ \Eprint
  {http://arxiv.org/abs/hep-ph/9310369} {arXiv:hep-ph/9310369} \BibitemShut
  {NoStop}%
\bibitem [{\citenamefont {Sharpe}\ and\ \citenamefont
  {Singleton}(1998)}]{Sharpe:1998xm}%
  \BibitemOpen
  \bibfield  {author} {\bibinfo {author} {\bibfnamefont {S.~R.}\ \bibnamefont
  {Sharpe}}\ and\ \bibinfo {author} {\bibfnamefont {R.~L.}\ \bibnamefont
  {Singleton}, \bibfnamefont {Jr}},\ }\href {\doibase
  10.1103/PhysRevD.58.074501} {\bibfield  {journal} {\bibinfo  {journal} {Phys.
  Rev. D}\ }\textbf {\bibinfo {volume} {58}},\ \bibinfo {pages} {074501}
  (\bibinfo {year} {1998})},\ \Eprint {http://arxiv.org/abs/hep-lat/9804028}
  {arXiv:hep-lat/9804028} \BibitemShut {NoStop}%
\bibitem [{\citenamefont {Rupak}\ and\ \citenamefont
  {Shoresh}(2002)}]{Rupak:2002sm}%
  \BibitemOpen
  \bibfield  {author} {\bibinfo {author} {\bibfnamefont {G.}~\bibnamefont
  {Rupak}}\ and\ \bibinfo {author} {\bibfnamefont {N.}~\bibnamefont
  {Shoresh}},\ }\href {\doibase 10.1103/PhysRevD.66.054503} {\bibfield
  {journal} {\bibinfo  {journal} {Phys. Rev. D}\ }\textbf {\bibinfo {volume}
  {66}},\ \bibinfo {pages} {054503} (\bibinfo {year} {2002})},\ \Eprint
  {http://arxiv.org/abs/hep-lat/0201019} {arXiv:hep-lat/0201019} \BibitemShut
  {NoStop}%
\bibitem [{\citenamefont {Aoki}\ \emph
  {et~al.}(2006{\natexlab{a}})\citenamefont {Aoki}, \citenamefont {Bar},
  \citenamefont {Takeda},\ and\ \citenamefont {Ishikawa}}]{Aoki:2005mb}%
  \BibitemOpen
  \bibfield  {author} {\bibinfo {author} {\bibfnamefont {S.}~\bibnamefont
  {Aoki}}, \bibinfo {author} {\bibfnamefont {O.}~\bibnamefont {Bar}}, \bibinfo
  {author} {\bibfnamefont {S.}~\bibnamefont {Takeda}}, \ and\ \bibinfo {author}
  {\bibfnamefont {T.}~\bibnamefont {Ishikawa}},\ }\href {\doibase
  10.1103/PhysRevD.73.014511} {\bibfield  {journal} {\bibinfo  {journal} {Phys.
  Rev. D}\ }\textbf {\bibinfo {volume} {73}},\ \bibinfo {pages} {014511}
  (\bibinfo {year} {2006}{\natexlab{a}})},\ \Eprint
  {http://arxiv.org/abs/hep-lat/0509049} {arXiv:hep-lat/0509049} \BibitemShut
  {NoStop}%
\bibitem [{\citenamefont {Aoki}\ \emph
  {et~al.}(2006{\natexlab{b}})\citenamefont {Aoki}, \citenamefont {Bar},\ and\
  \citenamefont {Takeda}}]{Aoki:2006ab}%
  \BibitemOpen
  \bibfield  {author} {\bibinfo {author} {\bibfnamefont {S.}~\bibnamefont
  {Aoki}}, \bibinfo {author} {\bibfnamefont {O.}~\bibnamefont {Bar}}, \ and\
  \bibinfo {author} {\bibfnamefont {S.}~\bibnamefont {Takeda}},\ }\href
  {\doibase 10.1103/PhysRevD.73.094501} {\bibfield  {journal} {\bibinfo
  {journal} {Phys. Rev. D}\ }\textbf {\bibinfo {volume} {73}},\ \bibinfo
  {pages} {094501} (\bibinfo {year} {2006}{\natexlab{b}})},\ \Eprint
  {http://arxiv.org/abs/hep-lat/0601019} {arXiv:hep-lat/0601019} \BibitemShut
  {NoStop}%
\bibitem [{\citenamefont {Lutz}\ \emph {et~al.}(2020)\citenamefont {Lutz},
  \citenamefont {Sauerwein},\ and\ \citenamefont {Timmermans}}]{Lutz:2020dfi}%
  \BibitemOpen
  \bibfield  {author} {\bibinfo {author} {\bibfnamefont {M.~F.~M.}\
  \bibnamefont {Lutz}}, \bibinfo {author} {\bibfnamefont {U.}~\bibnamefont
  {Sauerwein}}, \ and\ \bibinfo {author} {\bibfnamefont {R.~G.~E.}\
  \bibnamefont {Timmermans}},\ }\href {\doibase 10.1140/epjc/s10052-020-8417-5}
  {\bibfield  {journal} {\bibinfo  {journal} {Eur. Phys. J. C}\ }\textbf
  {\bibinfo {volume} {80}},\ \bibinfo {pages} {844} (\bibinfo {year} {2020})},\
  \Eprint {http://arxiv.org/abs/2003.10158} {arXiv:2003.10158 [hep-lat]}
  \BibitemShut {NoStop}%
\bibitem [{\citenamefont {Sauerwein}\ \emph {et~al.}(2022)\citenamefont
  {Sauerwein}, \citenamefont {Lutz},\ and\ \citenamefont
  {Timmermans}}]{Sauerwein:2021jxb}%
  \BibitemOpen
  \bibfield  {author} {\bibinfo {author} {\bibfnamefont {U.}~\bibnamefont
  {Sauerwein}}, \bibinfo {author} {\bibfnamefont {M.~F.~M.}\ \bibnamefont
  {Lutz}}, \ and\ \bibinfo {author} {\bibfnamefont {R.~G.~E.}\ \bibnamefont
  {Timmermans}},\ }\href {\doibase 10.1103/PhysRevD.105.054005} {\bibfield
  {journal} {\bibinfo  {journal} {Phys. Rev. D}\ }\textbf {\bibinfo {volume}
  {105}},\ \bibinfo {pages} {054005} (\bibinfo {year} {2022})},\ \Eprint
  {http://arxiv.org/abs/2105.06755} {arXiv:2105.06755 [hep-ph]} \BibitemShut
  {NoStop}%
\bibitem [{\citenamefont {Bulava}\ \emph {et~al.}(2022)\citenamefont {Bulava},
  \citenamefont {Hansen}, \citenamefont {Hansen}, \citenamefont {Patella},\
  and\ \citenamefont {Tantalo}}]{Bulava:2021fre}%
  \BibitemOpen
  \bibfield  {author} {\bibinfo {author} {\bibfnamefont {J.}~\bibnamefont
  {Bulava}}, \bibinfo {author} {\bibfnamefont {M.~T.}\ \bibnamefont {Hansen}},
  \bibinfo {author} {\bibfnamefont {M.~W.}\ \bibnamefont {Hansen}}, \bibinfo
  {author} {\bibfnamefont {A.}~\bibnamefont {Patella}}, \ and\ \bibinfo
  {author} {\bibfnamefont {N.}~\bibnamefont {Tantalo}},\ }\href {\doibase
  10.1007/JHEP07(2022)034} {\bibfield  {journal} {\bibinfo  {journal} {JHEP}\
  }\textbf {\bibinfo {volume} {07}},\ \bibinfo {pages} {034} (\bibinfo {year}
  {2022})},\ \Eprint {http://arxiv.org/abs/2111.12774} {arXiv:2111.12774
  [hep-lat]} \BibitemShut {NoStop}%
\bibitem [{\citenamefont {Ruiz~de Elvira}\ \emph {et~al.}(2018)\citenamefont
  {Ruiz~de Elvira}, \citenamefont {Hoferichter}, \citenamefont {Kubis},\ and\
  \citenamefont {Mei\ss{}ner}}]{RuizdeElvira:2017stg}%
  \BibitemOpen
  \bibfield  {author} {\bibinfo {author} {\bibfnamefont {J.}~\bibnamefont
  {Ruiz~de Elvira}}, \bibinfo {author} {\bibfnamefont {M.}~\bibnamefont
  {Hoferichter}}, \bibinfo {author} {\bibfnamefont {B.}~\bibnamefont {Kubis}},
  \ and\ \bibinfo {author} {\bibfnamefont {U.-G.}\ \bibnamefont
  {Mei\ss{}ner}},\ }\href {\doibase 10.1088/1361-6471/aa9422} {\bibfield
  {journal} {\bibinfo  {journal} {J. Phys. G}\ }\textbf {\bibinfo {volume}
  {45}},\ \bibinfo {pages} {024001} (\bibinfo {year} {2018})},\ \Eprint
  {http://arxiv.org/abs/1706.01465} {arXiv:1706.01465 [hep-ph]} \BibitemShut
  {NoStop}%
\bibitem [{\citenamefont {Gupta}\ \emph {et~al.}(2021)\citenamefont {Gupta},
  \citenamefont {Park}, \citenamefont {Hoferichter}, \citenamefont
  {Mereghetti}, \citenamefont {Yoon},\ and\ \citenamefont
  {Bhattacharya}}]{Gupta:2021ahb}%
  \BibitemOpen
  \bibfield  {author} {\bibinfo {author} {\bibfnamefont {R.}~\bibnamefont
  {Gupta}}, \bibinfo {author} {\bibfnamefont {S.}~\bibnamefont {Park}},
  \bibinfo {author} {\bibfnamefont {M.}~\bibnamefont {Hoferichter}}, \bibinfo
  {author} {\bibfnamefont {E.}~\bibnamefont {Mereghetti}}, \bibinfo {author}
  {\bibfnamefont {B.}~\bibnamefont {Yoon}}, \ and\ \bibinfo {author}
  {\bibfnamefont {T.}~\bibnamefont {Bhattacharya}},\ }\href {\doibase
  10.1103/PhysRevLett.127.242002} {\bibfield  {journal} {\bibinfo  {journal}
  {Phys. Rev. Lett.}\ }\textbf {\bibinfo {volume} {127}},\ \bibinfo {pages}
  {242002} (\bibinfo {year} {2021})},\ \Eprint
  {http://arxiv.org/abs/2105.12095} {arXiv:2105.12095 [hep-lat]} \BibitemShut
  {NoStop}%
\end{thebibliography}%
\bibliographystyle{apsrev4-1}
\end{document}